\documentclass[usenatbib]{mn2e}
\usepackage{natbibmnfix,graphicx,times}
\usepackage{epstopdf}
\usepackage{natbib}
\usepackage{hyperref}
\newcommand{\HI}{\mathrm{H\,I}}
\newcommand{\HII}{\mathrm{H\,II}}
\newcommand{\HeI}{\mathrm{He\,I}}
\newcommand{\HeII}{\mathrm{He\,II}}
\newcommand{\HeIII}{\mathrm{He\,III}}
\newcommand{\HIa}{H\,{\sevensize{\textbf{I}}}\,\,}
\newcommand{\HIb}{H\,{\sevensize{\textbf{I}}}}
\newcommand{\HeIa}{He\,\sevensize{\textbf{I}}\normalsize\,\,}
\newcommand{\HeIb}{He\,{\sevensize{\textbf{I}}}}
\newcommand{\HeIIa}{He\,{\sevensize{\textbf{II}}}\,\,}
\newcommand{\HeIIb}{He\,{\sevensize{\textbf{II}}}}

\newcommand{\GHI}{\Gamma_{\HI}}

\newcommand{\lya}{Ly$\alpha$ }
\newcommand{\lyam}{{\mathrm{Ly}\alpha}}
\newcommand{\IF}{\mathrm{IF}}
\newcommand{\Nion}{\dot{N}_\mathrm{ion}}

\newcommand{\apj}{ApJ}
\newcommand{\apjl}{ApJ}

\newcommand{\aj}{AJ}
\newcommand{\mnras}{MNRAS}

\newcommand{\physrep}{Physics Reports}
\newcommand{\prd}{PRD}

\newcommand{\nat}{Nature}

\newcommand{\pasp}{PASP}

\title[Quasar IF \lya Emission]{Quasar Ionization Front \lya Emission in an Inhomogeneous Intergalactic Medium} 
\author[F.~B. Davies, S.~R. Furlanetto, M. McQuinn]{Frederick B. Davies$^1$\thanks{davies@astro.ucla.edu}, Steven R. Furlanetto$^1$, Matthew McQuinn$^2$\\
$^1$Department of Physics \& Astronomy, University of California, Los Angeles, Box 951547, Los Angeles, CA 90095 \\
$^2$Department of Astronomy, University of California, Berkeley, CA 94720}

\begin{document}

\maketitle

\begin{abstract}
The conditions within the ionization front of a quasar during reionization ($T\sim30,000$ K, neutral hydrogen fraction $x_\HI\sim0.5$) are ideal for producing \lya emission via collisional excitation of hydrogen atoms. Observations of this emission, which could subtend  $\ga10$ arcmin$^2$ on the sky, would definitively demonstrate the presence of a neutral intergalactic medium at the observed epoch, placing valuable constraints on the progress of reionization. We find that the expected \lya surface brightness is significantly weaker than previously determined and may be impossible to observe with current and near-future instruments. Past work calculated the \lya emission from a quasar ionization front in a homogeneous medium with a clumping factor approximation to account for inhomogeneities. We find using 1D radiative transfer calculations that this approximation overestimates the emission by a factor of $\ga3$. Our calculations model the propagation of ionizing photons and compute the \lya emission from quasar ionization fronts on sightlines from a hydrodynamic cosmological simulation at $z=7.1$.  To better understand the physical properties of the emission, we also develop an analytic model that accurately describes the results of the full radiative transfer calculation.
\end{abstract}


\section{Introduction}\label{sec:intro}
Following the recombination of the universe probed by the cosmic microwave background (CMB) at $z\sim1100$, the universe consisted almost entirely of neutral hydrogen and helium. The first collapsed structures, stars, and galaxies proceeded to ionize the universe into the state that persists today \citep{stevebook}. The faint galaxies that produced the majority of these ionizing photons are currently out of reach of modern instruments \citep{Robertson13,Oesch13,Ellis13}, so the epoch of reionization is a valuable indirect probe of early structure formation. The time and duration of the reionization process is still under intense observational and theoretical investigation.

The simplest constraint on the epoch of reionization is the measurement of the electron scattering optical depth $\tau_\mathrm{e}$ by CMB experiments, which determines the average column density of electrons between the present day and the recombination epoch. Current measurements of $\tau_\mathrm{e}$ are consistent with an instantaneous reionization at $z\sim10.6$ \citep{Hinshaw13}, but modeling of the reionization process (e.g. \citealt{Furlanetto2004}) and kinetic Sunyaev-Zeldovich effect constraints \citep{Zahn12,Mesinger12} suggest that it should be extended over $3\la\Delta z\la7$. More model-dependent constraints have been obtained by studying \lya absorption in high-redshift quasar spectra, which suggest that the universe has been highly ionized since at least $z\sim6$ \citep{Fan2006}. Recently, an observed drop in the fraction of broadband color-selected $z\ga7$ galaxies that show bright \lya emission relative to $z\sim6$ has also been used to infer a substantial increase in the neutral fraction \citep{Stark10,Pentericci11,Treu12,Ono12}. Proposed measurements of 21 cm emission from the neutral cosmic web during reionization are a promising probe of the ionization state of the IGM \citep{Furlanetto2006}, especially in light of recent advances in foreground suppression (e.g. \citealt{Parsons14}), but detection of this emission still eludes current instruments. The search for a definitive probe of reionization that is accessible to current or near-future instruments continues.

\citet{C08} (henceforth C08) introduced a novel method to study the progression and topology of reionization:\footnote{The idea of observing large-scale \lya emission from the IGM during reionization was first investigated by \citet{Baltz98}, although in the different context of recombination emission from the entire IGM.} \lya emission from the ionization front (IF) of a luminous quasar. While quasars are unlikely to have been an important source of ionizing photons during the reionization epoch \citep{Fan2006}, their (rare) ionized bubbles will be the largest coherent structures in the universe during reionization. C08 showed that the conditions within quasar IFs are ideal for producing \lya emission. By definition, the IF is the narrow boundary between the neutral IGM (with a neutral fraction close to unity) and the inside of the ionized bubble (with a neutral fraction close to zero). Collisionally excited \lya emission is strongest when the number of hydrogen atoms equals the number of free electrons, which occurs inside the IF. Also, quasar ionizing spectra are harder than the typically assumed ionizing spectrum of galaxies, and the extra energy of the ionizing photons heats up the gas to 2--4$\times 10^4$ K. This temperature is hot enough that collisional excitations to the first excited state are efficient, and a substantial fraction of the absorbed energy is released as \lya photons. This \lya emission is analogous to \lya fluorescence of optically-thick IGM clouds at intermediate redshift \citep{GW96}, but instead of relying on the static ionized skin of a dense system, the emission can arise from any part of the (initially) neutral IGM as the IF passes through it. 

In their fiducial model, C08 calculated the ionization and heating of a uniform IGM with clumping factor $C\equiv\langle n^2 \rangle/\langle n \rangle^2=35$ by a quasar at $z=6.5$ and performed \lya radiative transfer to determine whether this \lya emission could be observable. The observed ionized region around a luminous quasar depends strongly on finite speed-of-light effects \citep{Yu05} which are computationally expensive to include in a radiative transfer code, but they corrected for this effect by re-scaling the timesteps in their simulation depending on the speed of IF expansion (we discuss this method further in Section \ref{sect:causal}). In the end, they found that the \lya emission would appear as faint, large-scale (a few proper Mpc across, $\sim$ several arcmin$^{2}$) line emission with \lya surface brightness $SB_\lyam \sim10^{-20} ((1+z)/6.5)^{-2}$ erg s$^{-1}$ cm$^{-2}$ arcsec$^{-2}$ around quasars with luminosities similar to the population known at $z\ga6$, which is just bright enough to be barely detectable with current instruments.

We improve on the C08 calculation by performing ionizing continuum radiative transfer through an inhomogeneous IGM. As we show in the next section, the overall clumping factor of the IGM is insufficient to describe the physics of density inhomogeneities in the IF. We also include secondary ionizations by high energy photoelectrons, which significantly modify the shape and temperature of the IF, and correct for causal effects when the IF is propagating close to the speed of light. The net result of these effects is a substantial decrease in the expected \lya emission, pushing it out of reach of existing instruments.

The structure of the paper is as follows. In Section \ref{sec:analyticdesc} we summarize a basic analytic model for IF \lya emission from an inhomogeneous IGM. In Section \ref{sect:numericalmethod} we describe our numerical methods including one-dimensional radiative transfer and correction for causal effects. In Section \ref{sect:rtresults} we describe the resulting ionization and temperature structure of our radiative transfer models in addition to the custom cosmological simulation from which we draw inhomogeneous IGM sightlines. In Section \ref{sect:analyticmethod} we introduce an analytic method, calibrated to our radiative transfer results, that allows rapid, accurate computation of the causal-corrected \lya emission along an IGM sightline and sheds light on the processes driving the IF \lya emission. Finally, in Section \ref{sect:lyaresults} we discuss the results of our \lya surface brightness calculation and investigate the detectability of this signal with current and upcoming instruments.

In this work we assume a $\Lambda$CDM cosmology with $\Omega_m=0.3$, $\Omega_\Lambda=0.7$, $\Omega_b=0.048$, and $h=0.68$.

\section{Ionization Front \lya Emission -- Analytic Description}\label{sec:analyticdesc}
To build intuition, in this section we describe the basic physics involved in the \lya emission from the IF. In particular, we discuss the dependence of the IF \lya emission on the local density, which has not been addressed in past work.

For the conditions within a quasar IF, $T\sim30,000$ K and neutral hydrogen fraction $x_\HI\sim0.5$, the collisional excitation rate of \lya can be orders of magnitude higher than the recombination rate within the ionized region (C08). To first order, the intensity of \lya emission from the IF is simply a function of two things: the width of the IF, $dR_\mathrm{IF}$, and the average \lya emissivity within the IF, $\langle\epsilon_\mathrm{Ly\alpha}\rangle$. The width of the IF is related to the mean free path of ionizing photons, or $dR_\mathrm{IF} \propto (n_\mathrm{H} \bar{\sigma}_\HI)^{-1}$, where $\bar{\sigma}_\HI$ is the effective ionization cross-section of ionizing photons within the IF and depends on the shape of the ionizing spectrum. The average energy of ionizing photons within the IF is a non-trivial function of the spectral index \citep{AH99}, but is mostly constant as a function of time. Thus, along a sightline, the width of the IF varies as $dR_\mathrm{IF}\propto n_\mathrm{H}^{-1}$. The \lya emissivity within the IF is a function of both density and temperature: $\epsilon_\mathrm{Ly\alpha} \propto n_\HI n_e q_\mathrm{eff}(T)$, where $q_\mathrm{eff}$ is the effective collisional excitation coefficient (C08). Assuming a constant temperature and ignoring the details of the $x_\HI$ profile within the IF, we can write $\langle \epsilon_\mathrm{Ly\alpha} \rangle\propto n_\mathrm{H}^2$. Finally, we arrive at the density dependence of the intensity of \lya emission for a uniform density medium: $I_\mathrm{Ly\alpha} \propto dR_\mathrm{IF} \langle\epsilon_\mathrm{Ly\alpha}\rangle \propto n_\mathrm{H}$. In a scenario where the material within the IF is at roughly a single density, the \lya emission is proportional to density instead of density squared, so the average intensity along a sightline will not necessarily reflect the overall clumping factor of the medium as assumed by C08.

The previous discussion assumed that the temperature within the IF was constant. However, this is not true in general. The temperature within the IF is largely a result of the competition between photoionization heating and (predominantly) collisional cooling processes. At a given radius from the source, the heating rate is only a function of the optical depth structure of the IF, which should be roughly constant with density, but the collisional cooling rate will scale as $n_\mathrm{H}^2$. A simple analytic scaling can be derived as follows: assume there is a fixed temperature to which the IF would heat the medium in the absence of cooling, $T_\mathrm{max}$, reflecting the energy deposited by photoheating. The presence of cooling will reduce this temperature by an amount $\Delta T \propto t_\mathrm{IF} L_\mathrm{cool} / N_\mathrm{particles}$, where $t_\mathrm{IF}$ is the time a parcel of gas is within the IF, $L_\mathrm{cool}$ is the rate of energy loss to cooling, and $N_\mathrm{particles}$ is the number of gas particles. The time within the IF can be written as $t_\mathrm{IF} = dR_\mathrm{IF} / v_\mathrm{IF}$ where $v_\mathrm{IF}$ is the \emph{instantaneous} velocity of the IF. The velocity of the IF is determined by the local flux of ionizing photons and density of neutral atoms. For a steady source and assuming constant density within the IF we have $v_\IF \propto n_\mathrm{H}^{-1} R^{-2}$. Because $dR_\IF \propto n_\mathrm{H}^{-1}$ as discussed before, we have $t_\IF \propto R^2$. If one further assumes a constant cooling rate within the IF dominated by \lya excitation, $L_\mathrm{cool} / N_\mathrm{particles} \propto \epsilon_\lyam / n_\mathrm{H} \propto n_\mathrm{H}$. Thus, $\Delta T \propto n_\mathrm{H} R^2$. As expected, denser regions cool more due to collisions, and as the IF slows down at large radii this effect becomes stronger. The effect will in general be much weaker than the scaling derived here because the \lya emissivity is a very strong function of temperature and will not remain constant. We derive a more accurate scaling of the IF temperature in Section \ref{sect:analytictemp}. The resulting lower temperature within dense regions further weakens the density scaling of collisional \lya emission from the IF.

The next step is to consider the inhomogeneous IGM. One way to account for inhomogeneities is to enhance the rate of collisional processes by a clumping factor $C=\langle n^2 \rangle/\langle n \rangle^2$ which can be estimated by cosmological simulations.
This approximation assumes that the region of interest covers a broad range of densities representative of the IGM as a whole. This is likely reasonable for the large-scale roughly spherical extent of the IF. However, along the line of sight to the ionizing source, the IF is a relatively narrow structure: $\sim15$ proper kpc in width at the mean density of the universe at $z=7$ for a power law ionizing spectrum typical of luminous quasars\footnote{The IF is considerably wider than the mean free path at the ionizing edge ($\bar{n}_\mathrm{H}\sigma_\HI^{-1}\sim1$ kpc at $z=7$) because the average energy of ionizing photons within the IF is $\sim3\nu_\HI$.}. The intensity of \lya emission from the IF depends on this line of sight profile: when the IF passes through a dense region, it becomes narrower, and as discussed above the temperature of the gas may also change. We will show that \emph{the IF resolves clumping in the IGM}. This means that a clumping factor approximation overestimates the enhancement due to overdense regions. Instead, the average emission from a wider area that more broadly samples the density field will depend on the relative amount of time spent in low versus high density regions, as well as other effects that we will discuss in Section \ref{sect:lyaresults}. The amount of time that the IF spends in a given overdense region of size $dr$ will be $t_\IF \sim dr/v_\IF \propto n_\mathrm{H} \, dr $. If one assumes (as a simple toy model) that $dr$ scales with the Jeans length of the gas, then $dr \propto n_\mathrm{H}^{-1/2}$ and so $t_\IF \propto n_\mathrm{H}^{1/2}$. Thus, the IF would spend more time in overdense regions, but not enough to recover the $n_\mathrm{H}^2$ scaling assumed by a clumping factor.

\section{Numerical Method}\label{sect:numericalmethod}
\subsection{1D radiative transfer}\label{sect:rtdesc}
To calculate the time-dependent properties of the IF, we developed a one-dimensional radiative transfer model based on the method of \citet{BH07}. We assume a medium consisting solely of hydrogen and helium at their primordial ratios and a single steady source of ionizing radiation. The radiative transfer model solves the following time-dependent equations governing the abundance of ionized species as a function of time and distance from the ionizing source:
\begin{eqnarray}
\frac{dn_\HII}{dt} &=& n_\HI \Gamma_\HI - n_\HII n_e \alpha^A_\HII, \\
\frac{dn_\HeII}{dt} &=& n_\HeI \Gamma_\HeI + n_\HeIII n_e \alpha^A_\HeIII \nonumber \\
&& - n_\HeII (\Gamma_\HeII - n_e \alpha^A_\HeII), \\
\frac{dn_\HeIII}{dt} &=& n_\HeII \Gamma_\HeII - n_\HeIII n_e \alpha^A_\HeIII, \\
\end{eqnarray}
where $n_i$, $\Gamma_i$, and $\alpha^A_i$ are the number densities, ionization rates, and Case A recombination rate coefficients\footnote{Case A is a reasonable approximation to a full treatment of recombination photons when $t \ll t_\mathrm{rec}$ (see, e.g., discussion in \citealt{CP11})} \citep{HG97}, respectively. The remaining species are solved by the closing conditions
\begin{eqnarray}
n_\HI &=& n_\mathrm{H} - n_\HII, \\
n_\HeI &=& \frac{Y}{4(1-Y)}n_\mathrm{H} - n_\HeII - n_\HeIII, \\
n_e &=& n_\HII + n_\HeII + 2n_\HeIII,
\end{eqnarray}
where $Y=0.24$ is the mass fraction of helium.

The ionization rate of species $i$, $\Gamma_i$, consists of photoionization by the central source (or ``primary" ionization), ``secondary" ionization by energetic photoelectrons \citep{Shull85}, and collisional ionization by thermal electrons \citep{Theuns98}. The primary photoionization rate is given by
\begin{equation}\label{eqn:firstion}
n_i \Gamma^{\gamma}_{i,1} = \frac{1}{dV} \int_{\nu_i}^\infty \frac{L_\nu e^{-\tau_\nu}}{h\nu} P_i d\nu,
\end{equation}
where $L_\nu\propto\nu^{-\alpha_Q}$ is the specific luminosity of the ionizing source at frequency $\nu$ with spectral index $\alpha_Q$, $dV$ is the volume of a spherical shell of width $dR$ at distance $R$ from the source, and $\tau_\nu$ is the total optical depth to photons of frequency $\nu$ at distance $R$ from the source. $P_i$ is the probability that species $i$ is ionized by a photon with frequency $\nu$ given by \citep{Bolton04}
\begin{eqnarray}
P_\HI &=& p_\HI q_\HeI q_\HeII (1-\mathrm{e}^{-\tau_\nu^\mathrm{tot}})/D, \\
P_\HeI &=& q_\HI p_\HeI q_\HeII (1-\mathrm{e}^{-\tau_\nu^\mathrm{tot}})/D, \\
P_\HeII &=& q_\HI q_\HeI p_\HeII (1-\mathrm{e}^{-\tau_\nu^\mathrm{tot}})/D,
\end{eqnarray}
where $p_i=1-\mathrm{e}^{-\tau_\nu^i}$, $q_i = \mathrm{e}^{-\tau_\nu^i}$, $\tau_\nu^i = n_i \sigma_i(\nu) dR$ is the optical depth of species $i$ in a given cell using photoionization cross-sections $\sigma_i(\nu)$ from \citet{Verner96},  
$\tau_\nu^\mathrm{tot} = \tau_\nu^\HI + \tau_\nu^\HeI + \tau_\nu^\HeII$, and $D = p_\HI q_\HeI q_\HeI + q_\HI p_\HeI q_\HeII + q_\HI q_\HeI p_\HeII$.

We include secondary ionizations by energetic photoelectrons as a function of electron energy using the results of \citet{FJS10} (henceforth FJS10). The secondary ionization rate is given by
\begin{equation}\label{eqn:secondion}
n_i \Gamma^{\gamma}_{i,2} = \frac{1}{dV}\int_{\nu_i}^\infty f^\mathrm{ion}_{i,\nu} \left(\frac{h\nu-h\nu_i}{h\nu_i}\right) \frac{L_\nu e^{-\tau_\nu}}{h\nu} P_i d\nu,
\end{equation}
where $f^\mathrm{ion}_{i,\nu}$, the fraction of photoelectron energy that goes into secondary ionization of species $i$, is calculated by interpolating the publicly available tables of FJS10.

Much of the remaining photoelectron energy is converted into thermal energy in the gas through collisions, with photoheating rates $\epsilon_i$ given by
\begin{equation}\label{eqn:heating}
n_i \epsilon_i = \frac{1}{dV}\int_{\nu_i}^\infty f^\mathrm{heat}_{i,\nu} (h\nu-h\nu_i) \frac{L_\nu e^{-\tau_\nu}}{h\nu} P_i d\nu,
\end{equation}
where $f^\mathrm{heat}_{i,\nu}$ is the fraction of photoelectron energy that goes into heating the gas after a photon of frequency $\nu$ ionizes species $i$. The remainder of the photoelectron energy is released as \lya emission, but this emission is unimportant compared to collisionally excited \lya that we are interested in.

The calculations by FJS10 assume that all of the interactions of the energetic photoelectrons occur instantaneously. However, this is not actually the case. The timescale over which an electron with energy $E$ is depleted by collisional ionization of hydrogen is roughly (FJS10)
\begin{equation}
t_{\mathrm{loss,H}} \sim 10^6 x_\HI^{-1} \left(\frac{E}{1\,\mathrm{keV}}\right)^{3/2} \Delta^{-1} \left( \frac{1+z}{8} \right) ^{-3} \mathrm{yr},
\end{equation}
where $\Delta = n_\mathrm{H}/\bar{n}_\mathrm{H}(z)$ is the density in units of the cosmic mean at redshift $z$. We include this timescale in an approximate manner by suppressing the secondary ionization rate in a given grid cell for $\sim t_{\mathrm{loss,H}}$ before the IF reaches the cell. To ensure conservation of energy, the energy that would have gone into secondary ionizations is instead directed into heating the gas. We multiply the fraction of energy going into secondary ionizations $f^\mathrm{ion}_{\HI,\nu}$ by $1-\mathrm{e}^{(t-t_\mathrm{IF})/t_\mathrm{loss}}$, where $t_\mathrm{IF}$ is an estimate of the time at which the IF will reach the current cell and $t$ is the current time in the simulation. The energy removed from secondary ionizations is then put into heating by adjusting $f^\mathrm{heat}_{\HI,\nu}$ accordingly. This approximation causes the temperature and ionization structure to match the no-secondaries model at very early times ($t \la 10^5$ yr). Over the next few Myr it converges to a model following the unadjusted FJS10 rates. A more precise calculation of this effect taking into account the evolution of the time-dependent energy spectrum of electrons would be ideal but is outside the scope of the present work.

The temperature evolution is determined by
\begin{equation}
\frac{dT}{dt} = \frac{(\gamma-1)\mu m_\mathrm{H}}{k_B \rho} (\mathcal{H}_\mathrm{tot} - \Lambda_\mathrm{tot}) - 2H(z)T - \frac{T}{n}\frac{dn_e}{dt},
\end{equation}
where $n = n_e + n_\mathrm{H} + n_\mathrm{He}$ is the total number density of all species, $\mathcal{H}_\mathrm{tot} = \sum_i n_i\epsilon_i$ is the total heating rate, $\mu$ is the mean molecular weight, $\Lambda_\mathrm{tot}$ is the total cooling rate, and $H(z)$ is the Hubble parameter. The cooling rate $\Lambda_\mathrm{tot}$ contains contributions from recombination cooling from \citet{HG97}, collisional excitation cooling and free-free emission from \citet{Cen92}, and inverse Compton cooling.

We compute the \lya emissivity in a similar manner to C08. The effective \lya collisional excitation coefficient is the sum over all collisional excitations from the ground state that lead to emission of a \lya photon, 
\begin{equation}
q_\mathrm{eff} = \sum_{n,l} f_{\lyam,nl} q_{1,nl},
\end{equation}
where $q_{1,nl}$ is the collisional excitation coefficient for the transition from the ground state to atomic level $nl$ and $f_{\lyam,nl}$ is the fraction of transitions back to the ground state from atomic level $nl$ that result in the emission of a \lya photon \citep{PF06}. We use the updated fits to the collisional excitation rate coefficients of \citet{Giovanardi87} from \citet{GP89} and include excitations up to $n=4$.

The calculation is discretized into a series of spatial grid cells that, for computational simplicity, correspond to spherical shells around the source. The physical structure of the IF depends on the density field, so to resolve the IF equally well at all times, we define the resolution of the spatial grid in terms of a hydrogen column density $N_\mathrm{H,cell}$. We find that adequate convergence of the \lya emission and temperature structure is achieved when $N_\mathrm{H,cell} \sim 1.2\times10^{18}$ cm$^{-2}$, corresponding to a spatial resolution of $dR \sim 4$ kpc $\Delta^{-1} ((1+z)/8)^{-3}$. In order to avoid wiping out small-scale fluctuations in the density field inside of voids, we require that $dR\geq dR(\Delta=1)$, so $N_\mathrm{H,cell}$ will vary somewhat in underdense regions. This converged resolution may seem rather coarse -- the initial optical depth at the hydrogen ionizing edge across each cell is $\sim10$. However, the radiative transfer algorithm we use is well-suited to such optically thick cells \citep{Bolton04} and we confirmed that the temperature and ionization structure have converged.

The integrals over frequency in equations (\ref{eqn:firstion}--\ref{eqn:heating}) are computed as a discrete sum over 80 logarithmic frequency bins\footnote{We did not attempt to optimize the number of frequency bins (e.g. \citealt{Mirocha12}) but this appears to be adequate for convergence.} from $\nu_\mathrm{ion,i}$ to $40\nu_\mathrm{ion,i}$ for each neutral (or partially neutral) species $i$ (\HIb, \HeIb, \HeIIb). In this sense we do not explicitly follow a single spectrum of photons but instead treat the ionizing spectrum of each species independently (except for the optical depths $\tau_\nu$ and absorption probabilities $P_i$ which include all species). 

The global time step $\Delta t$ is set by the speed of the IF. We require that the IF take more than one time step (typically $\ga 2$) to cross the current grid cell of the IF,
\begin{equation}
\Delta t = \frac{4\pi R_\IF^2 N_\mathrm{H,cell}}{\Nion},
\end{equation}
where $R_\IF$ is defined to be the first cell from the origin with $x_\HI > 0.5$. However, while $\Delta t$ is typically smaller than the light-travel time across the cell, it may be too coarse to accurately integrate the temperature and ionization state equations within the IF, so we loop over each cell with a sub-time step defined by requiring $\Delta n_i/n_i < 0.05$ for each species $i$ and $\Delta T/T < 0.05$. We compute the average transmission of ionizing photons through the cell during the sub-time step loop to propagate a time-averaged spectrum of ionizing photons to the next cell. With these strict criteria on the global and sub-time steps, we avoid numerical artifacts (such as, e.g., the temperature ``ringing" seen in \citealt{VB11}) and accurately compute ionization, heating, and cooling within the rapidly evolving IF.

\subsection{Causal correction}\label{sect:causal}
The numerical radiative transfer method described in the preceding section assumes an infinite speed of light for ease of calculation. This assumption can lead to unphysical effects -- most importantly IF velocities greater than the speed of light. Previous studies have found that the infinite speed of light calculation is an exact description of the IF propagation when observed along the line of sight to the source, and that the rest-frame behavior can be recovered by a simple change of coordinates (e.g. \citealt{White03,Shapiro06,BH07}). We discuss the detailed application of this effect to the IF below.

Let us assume a homogeneous medium for simplicity and an IF with a width $dR_\IF$ at radius $R_\IF(t)$ in the rest frame of the quasar. The IF will have a \lya emissivity profile
\begin{equation}
\epsilon_\lyam(R) = I_\mathrm{rest} F\left(\frac{R-R_\IF(t)}{dR_\IF}\right),
\end{equation}
where $F$ is a function describing the emission profile of the IF as a function of radius, normalized such that $\int F(R/dR_\IF) dR = 1$, and $I_\mathrm{rest}$ is the radially integrated emissivity of the IF. We ignore the time dependence of $I_\mathrm{rest}$ and $dR_\IF$ because the relevant timescale in the following is the light-crossing time of the IF. 

The \lya emission from the IF is observed on the light cone at angle $\theta$ from the line of sight (see Figure 3 of C08). The intensity on the light cone $I_\mathrm{LC}$ at time $t_\mathrm{LC}$ is then
\begin{eqnarray}\label{eqn:lightcone}
I_\mathrm{LC} &=& \int I_\mathrm{rest} F\left(\frac{R-R_\IF(t_\mathrm{LC}+R\cos{\theta}/c)}{dR_\IF}\right) dR \nonumber \\
&=& \int \left(1-\frac{v_\IF(t)}{c}\cos{\theta}\right)^{-1}I_\mathrm{rest} F\left(\frac{y}{dR_\IF}\right) dy \nonumber \\
&\approx& \left(1-\frac{v_\IF(t)}{c}\cos{\theta}\right)^{-1} I_\mathrm{rest},
\end{eqnarray}
where in the second line we have changed variables to $y=R-R_\IF(t_\mathrm{LC}+R\cos{\theta}/c)$ and in the third line we make the approximation that $v_\IF$ does not change over a light-crossing time. The time in the quasar rest-frame is $t=t_\mathrm{LC}+R\cos{\theta}/c$. Here, $v_\IF$ is the ``correct" IF velocity accounting for a finite speed of light ($v_\IF < c$). Thus, the observed intensity of the IF is corrected by the factor $[1-(v_\IF(t)/c)\cos{\theta}]^{-1}$. This factor is equivalent to the extra fractional time that a photon spends inside a moving IF relative to the static case ($v_\IF = 0$).

The infinite speed of light calculation provides IF \lya intensity $I_\mathrm{code}$ and IF velocity $v_{\IF,c=\infty}$ on the light cone for an observer at $\theta = 0$. This makes sense conceptually: in this frame, the order of photons as they are absorbed is the same as the order they were emitted, even for photons absorbed at much different distances. To convert this to any arbitrary $\theta$, we first rescale the time coordinate
\begin{eqnarray}\label{eqn:caustime}
t_\mathrm{code} + R/c = t_\mathrm{LC} + R \cos{\theta}/c \nonumber \\
t_\mathrm{LC} = t_\mathrm{code} + \frac{R}{c}(1-\cos{\theta})
\end{eqnarray}
at every $R$. In practice, one can instead implement this at $R=R_\IF$ by re-scaling the numerical time steps,
\begin{equation}\label{eqn:dtconversion}
dt_\mathrm{LC} = dt_\mathrm{code} \left(1+\frac{v_{\IF,c=\infty}}{c}(1-\cos{\theta})\right),
\end{equation}
as in C08. The intensity computed from the code is related to the rest-frame intensity through equation (\ref{eqn:lightcone}),
\begin{equation}
I_\mathrm{code} = (1-v_\IF/c)^{-1} I_\mathrm{rest},
\end{equation}
so the intensity on the light cone is then
\begin{eqnarray}
I_\mathrm{LC} &=& \left(1-\frac{v_\IF}{c}\cos{\theta}\right)^{-1} I_\mathrm{rest} \nonumber \\
&=& \frac{1-v_\IF/c}{1-(v_\IF/c)\cos{\theta}} I_\mathrm{code}.
\end{eqnarray}
The rest-frame IF velocity $v_\IF$ is related to the infinite speed of light IF velocity $v_{\IF,c=\infty}$ from the code by \citep{Shapiro06}
\begin{equation}
v_\IF = \frac{v_{\IF,c=\infty}}{1+v_{\IF,c=\infty}/c},
\end{equation}
so that
\begin{equation}\label{eqn:sbconversion}
I_\mathrm{LC} = \left(1 + \frac{v_{\IF,c=\infty}}{c}(1-\cos{\theta})\right)^{-1} I_\mathrm{code}.
\end{equation}

In order to properly compute the IF \lya emission by correcting the infinite speed of light calculation in the manner described above, the calculation must provide an accurate gas temperature. We show below that this is indeed the case.

An arbitrary gas property $Z$ at position $x$ along a sightline is only influenced by properties at $x'>x$ if $x'$ is on its backward light cone. We can write an equation describing the evolution of $Z$ with time,
\begin{equation}
Z(x,t) = Z(x,t-dt) + \int_x^\infty f(x',t-[x'-x]/c)dt dx',
\end{equation}
where $f$ is an unspecified function. If, for example, $Z$ is the ionization state of the gas, $f$ describes the absorption of photons at positions $x'$ along the light cone and the luminosity of the quasar at time $t-[x_\mathrm{Q}-x]/c$. Writing the above as a differential equation we find
\begin{equation}
\frac{dZ(x,t-x/c)}{dt} = \int_x^\infty dx' f(x',t-x'/c).
\end{equation}
Then, evaluating $Z$ on the light cone, 
\begin{equation}
\frac{dZ(x,t_\mathrm{LC})}{dt_\mathrm{LC}} = \int_x^\infty dx' f(x',t_\mathrm{LC}),
\end{equation}
using $t_\mathrm{LC} = t-x'/c$ and the fact that the Jacobian for the coordinate transformation $dtdx \rightarrow dt_\mathrm{LC}dx$ is equal to unity.

The equation above is identical to the equation that the infinite speed of light code solves for property $Z$:
\begin{equation}
\frac{dZ(x,t)}{dt} = \int_x^\infty dx' f(x',t).
\end{equation}
Thus, as long as the boundary conditions for $Z(x,t)$ are the same as the light cone time boundary conditions, the infinite speed of light calculation gives the same solution for the state of the gas as explicitly solving on the light cone.

While C08 corrected their infinite speed of light simulations to the time observed on the light cone with equation (\ref{eqn:dtconversion}), it appears they did not include the additional correction to the intensity in equation (\ref{eqn:sbconversion}). The physical reason for this correction is a difference in the width of the IF when a finite speed of light is taken into account. Consider the rest-frame ($\theta=\pi/2$) expansion of the IF. The width of a slow ($v_\IF \ll c$) IF is given by the optical depth of ionizing photons into the neutral medium, but for a fast ($v_\IF \sim c$) IF, the width is instead limited by the ionization timescale $t_\mathrm{ion} \sim \GHI^{-1}$ as neutral gas is illuminated inside the causal boundary $R=ct$. \citet{MR94} showed that the difference between these two widths is given by the correction factor in equation (\ref{eqn:sbconversion}) with $\theta=\pi/2$. For different $\theta$ the correction is analogous to having a finite speed of light equal to $c(1-\cos{\theta})^{-1}$ which is the maximum IF velocity along the light cone.

A full application of this correction requires changing coordinates via equation (\ref{eqn:caustime}) at every cell and timestep in the radiative transfer model. This requires significantly higher spatial and temporal resolution than the infinite speed of light calculation, so we do not apply the correction factor directly to the radiative transfer results presented in the following section. Instead, we apply the correction factor to a simplified, yet accurate, analytic model of IF \lya emission discussed in Section \ref{sect:analyticmethod}. In the rest of the paper, we will refer to the corrections to the time coordinate and \lya intensity as the ``causal correction".

\section{Radiative Transfer Results}\label{sect:rtresults}
In this section we present the results of our radiative transfer modeling for both a uniform IGM test case and for sightlines through an inhomogeneous IGM drawn from numerical simulations at $z=7.1$. The results presented in this section assume an infinite speed of light, and do not include the causal correction discussed in Section \ref{sect:causal}.

\begin{figure}
\begin{center}
\resizebox{8cm}{!}{\includegraphics{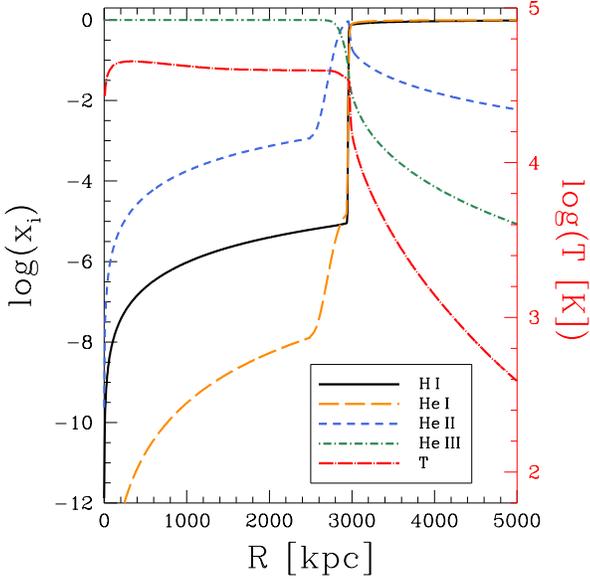}}\\
\end{center}
\caption{Ionization and temperature structure from the radiative transfer model in a uniform medium at $z=7$ with our fiducial uniform model parameters: $\Nion = 10^{57}$ s$^{-1}$, $\alpha_\mathrm{Q}=1.5$, $t_\mathrm{code}=10$ Myr. Species fractions $x_\HI$ (solid, black), $x_\HeI$ (long-dashed, orange), $x_\HeII$ (short-dashed, blue), and $x_\HeIII$ (dot-dashed, green) are shown along with the gas temperature (long-dot-dashed, red).}
\label{fig:unifall}
\end{figure}

\begin{figure}
\begin{center}
\resizebox{8cm}{!}{\includegraphics{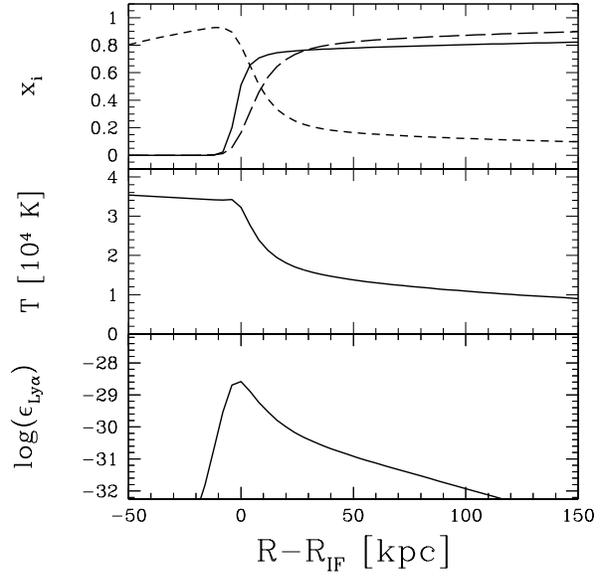}}\\
\end{center}
\caption{Zoom in on the IF ionization structure (top panel), temperature (middle), and \lya emissivity (bottom) of the fiducial uniform model (see Figure \ref{fig:unifall}). In the top panel the solid, long-dashed, and short-dashed curves are the species fractions $x_\HI$, $x_\HeI$, and $x_\HeII$, respectively.}
\label{fig:unifzoom}
\end{figure}

\begin{figure}
\begin{center}
\resizebox{8cm}{!}{\includegraphics{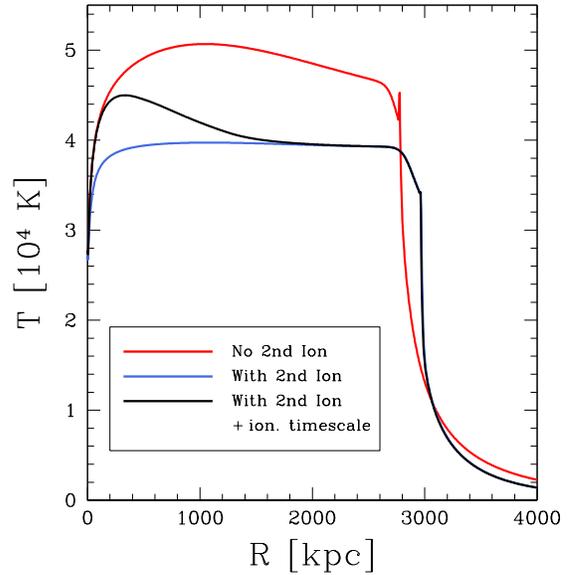}}\\
\end{center}
\caption{Dependence of the temperature structure on secondary ionizations and the secondary ionization timescale correction in the fiducial uniform model (see Figure \ref{fig:unifall}). The black curve is the fiducial prescription, the blue curve does not have the secondary ionization timescale correction (see text), and the red curve does not include secondary ionizations at all.}
\label{fig:secondtemp}
\end{figure}

\begin{figure}
\begin{center}
\resizebox{8cm}{!}{\includegraphics{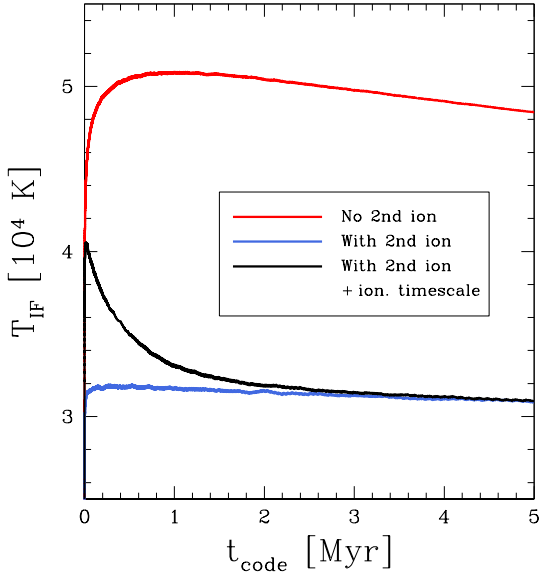}}\\
\end{center}
\caption{IF temperature as a function of time for models without secondary ionizations (red curve), with secondary ionizations (blue curve), and with secondary ionizations plus an ionization timescale correction (black curve) using the fiducial uniform model of Figure \ref{fig:unifall}.}
\label{fig:secondiftemp}
\end{figure}

\begin{figure}
\begin{center}
\resizebox{8cm}{!}{\includegraphics{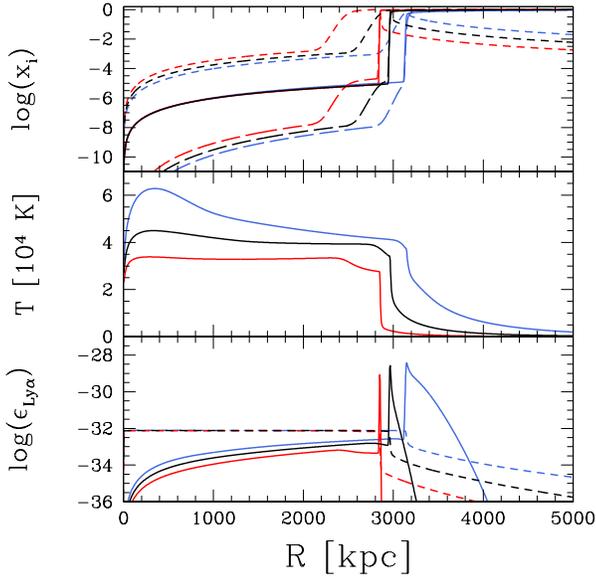}}\\
\end{center}
\caption{IF ionization structure (top), temperature (middle), and \lya emissivity of the fiducial uniform model (see Figure \ref{fig:unifall}) with varying quasar spectral index $\alpha_\mathrm{Q} =1.0,1.5,2.0$ (blue, black, red, respectively). In the top panel, the solid, long-dashed, and short-dashed curves are the species fractions $x_\HI$, $x_\HeI$, and $x_\HeII$, respectively. In the bottom panel, the solid curves are the \lya emissivity due to collisional excitation and the short-dashed curves are the \lya emissivity from recombinations.}
\label{fig:unifalpha}
\end{figure}

\begin{figure}
\begin{center}
\resizebox{8cm}{!}{\includegraphics{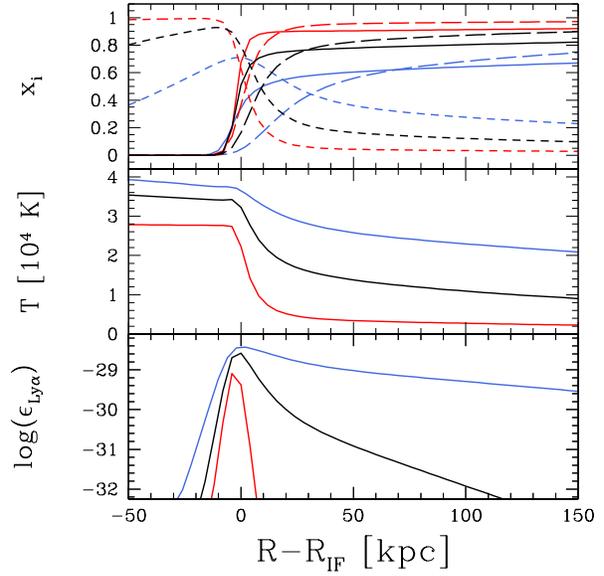}}\\
\end{center}
\caption{IF ionization structure (top), temperature (middle), and \lya emissivity (bottom) of the fiducial uniform model (see Figure \ref{fig:unifall}) with varying quasar spectral index $\alpha_\mathrm{Q}=1.0,1.5,2.0$ (blue, black, red, respectively). The line styles are the same as Figure \ref{fig:unifalpha}. Each curve has been shifted to $R - R_\IF$ for ease of comparison. }
\label{fig:unififalpha}
\end{figure}

\begin{figure}
\begin{center}
\resizebox{8cm}{!}{\includegraphics{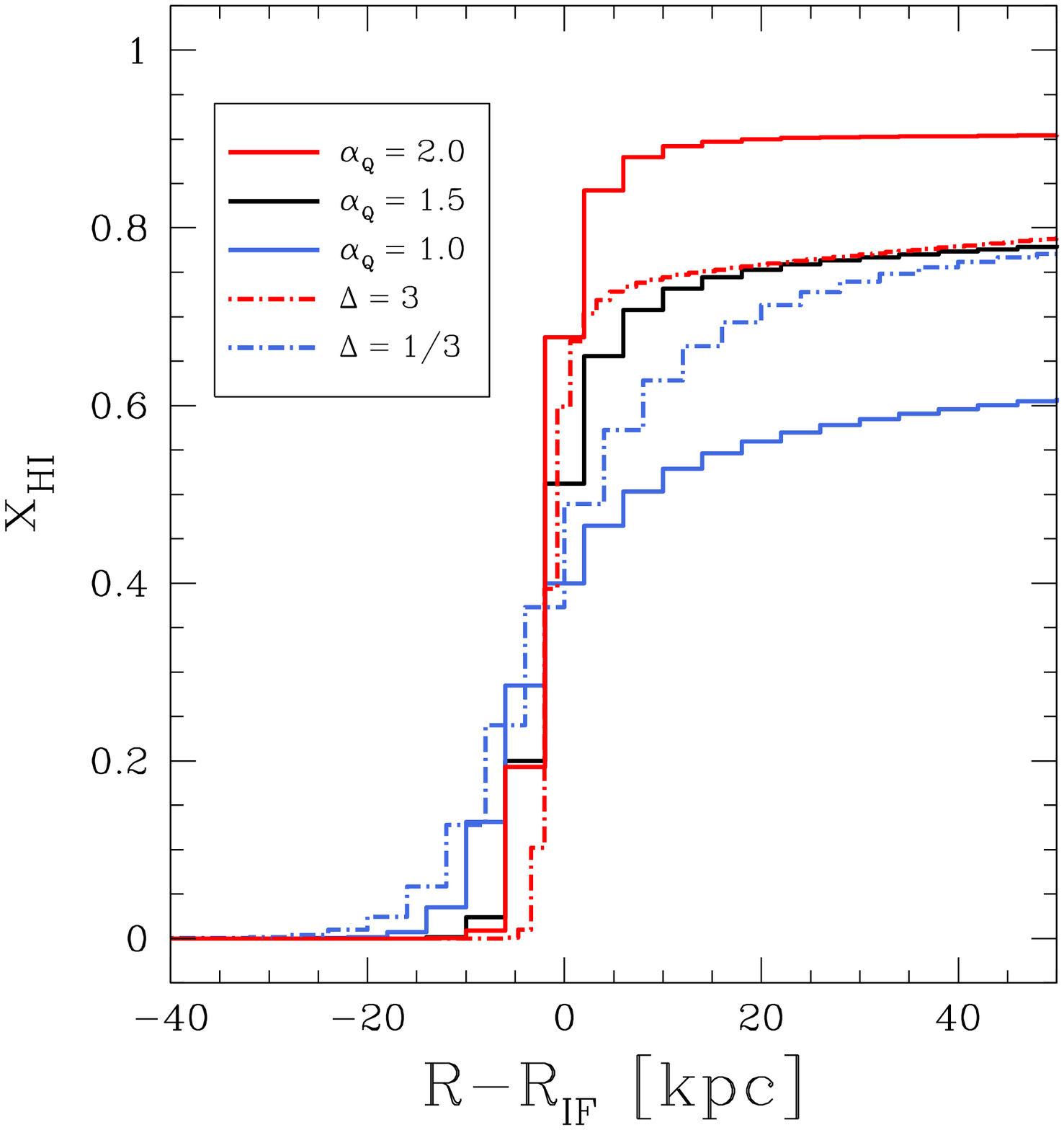}}\\
\end{center}
\caption{IF neutral hydrogen structure when the density (dot-dashed) and spectral index (solid) are varied to higher (red) and lower (blue) values compared to the fiducial uniform model of Figure \ref{fig:unifall} (black).}
\label{fig:unififall}
\end{figure}

\subsection{Uniform IGM}\label{sect:unifmodel}
We have run a series of simple models assuming a constant density IGM to demonstrate the response of the ionization and temperature structure to the treatment of secondary ionizations and source properties.
Our fiducial test model assumes a $\dot{N}_\mathrm{ion} = 10^{57}$ s$^{-1}$ quasar with EUV spectral index $\alpha=1.5$, similar to the inferred ionizing emission properties of known luminous $z\ga6$ quasars \citep{BH07,Mortlock11}, emitting into a uniform IGM with density equal to the cosmic mean at $z=7$ and an initial temperature of 10 K for $t_\mathrm{code}=10$ Myr. 

In Figure \ref{fig:unifall} we show the resulting ionized and neutral fractions for hydrogen and helium as well as the temperature structure. At large radii, hard UV/X-ray photons significantly ``preheat" the medium up to $\sim2\times10^4$ K and introduce a low level of ionization to the initially neutral medium ($x_\HII\sim0.05$). Figure \ref{fig:unifzoom} zooms in on the region close to the IF and shows that just beyond the IF lies a ``pre-ionization" region where secondary ionizations from high energy photons ionize $\sim10\%$ of the \HIa in a long tail outside of the IF. Within the IF itself, despite ongoing \HIa and \HeIa photoheating, the temperature stops increasing once $x_\HI\sim0.5$ -- this is when the cooling rate due to collisional excitation is strongest and the ionizations are predominantly due to the lowest energy ionizing photons. Within the \HIb/\HeIa ionized region lies the relatively broad \HeIIa IF which increases the gas temperature by another $\sim10^4$ K. The bottom panel of Figure \ref{fig:unifzoom} shows that the \lya emission is strongly peaked at the center of the IF, as expected, with a weak tail within the pre-ionization region. Assuming a (somewhat unrealistic) spherical shell morphology, the total integrated \lya luminosity from collisions within the IF is orders of magnitude larger than the recombination emission within the ionized bubble, assuming a uniform medium.

\subsubsection{Secondary Ionizations}\label{sect:secondion}
The inclusion of secondary ionizations has a substantial impact on the ionization and temperature structure of our radiative transfer model. Figure \ref{fig:secondtemp} shows the effect of our secondary ionization prescription on the temperature profile. A significant amount of energy that would otherwise go into heating the IGM instead goes into ionizations, causing a tradeoff between the size of the ionized bubble and the post-IF temperature. Our rough correction for the timescale of secondary ionizations causes the temperature profile to follow the no-secondaries model at early times, but Figure \ref{fig:secondiftemp} shows that the temperature within the IF is largely unaffected after $\sim2$ Myr. The IF is somewhat broadened by secondary ionizations because they increase the number of ionizations from higher energy photons that can penetrate farther into the neutral medium. Despite the broader IF, the substantially lower temperature results in a factor of a few less \lya emission.

\subsubsection{Variation of input parameters}\label{sect:unifparams}
Figure \ref{fig:unifalpha} shows the effect of changing the input quasar spectrum while leaving the total number of ionizing photons constant. The average energy of the ionizing photons regulates the enhanced ionization rate due to secondary ionizations, the heating of the IGM, and the width of the IF. The combination of the latter two effects causes the \lya emission to vary strongly with the assumed quasar spectrum. Figure \ref{fig:unififalpha} shows a zoom-in on the IF for $\alpha=1.0,1.5,2.0$. The enhanced secondary ionizations of the $\alpha=1.0$ model lead to an extended pre-ionization region. This pre-ionization is clear in Figure \ref{fig:unififall} where we show a zoom-in of $x_\HI$ around the IF for varying density and spectral index. The profile has an obvious ``knee" feature with more gradual increase in $x_\HI$ at further distances. Because the IF \lya emission is $\propto x_\HI(1-x_\HI)$, the stronger pre-ionization by hard ionizing spectra greatly increases the total emission. Note that the models with the same spectral index have a similar level of pre-ionization, suggesting that the effect is not limited by recombinations.

\subsection{Inhomogeneous IGM}\label{sect:inhomogmodel}
\subsubsection{Cosmological simulation}\label{sect:cosmosim}
To generate an inhomogeneous IGM density field, we ran a {\small GADGET-3} simulation \citep{Springel05} with a volume 12.5 Mpc$/h$ on a side with $512^3$ dark matter and gas particles to $z=7.1$, the redshift of the most distant quasar published to date \citep{Mortlock11}. The cosmological parameters were the same as assumed in the radiative transfer simulations: $\Omega_m=0.3$, $\Omega_\Lambda=0.7$, $\Omega_b=0.048$, $h=0.68$, and $\sigma_8=0.82$. To maximize clumping in the IGM so as to emphasize the effect of inhomogeneities on IF \lya emission, the simulation was run without photoheating by a uniform ionizing background. For computational efficiency a temperature floor of $500$ K was applied, which is compatible with some models for early heating of the neutral IGM by X-ray sources (e.g. \citealt{FurlanettoXray}, but see \citealt{Fialkov14}). We then drew 100 sightlines in random directions from the most massive halo in the box with $M_h\sim10^{11} M_\odot$. While this is likely at least an order of magnitude smaller than estimated halo masses for luminous high-redshift quasars ($M_h\sim10^{12-13} M_\odot$; \citealt{Walter04,Fan04,Willott05,Fanidakis13}), the scale of the local overdensity due to such a halo is very small compared to the size scales of the ionized region after just a few Myr and thus should not significantly impact our results. In any case, we regard properly resolving the IGM structure at the small scales relevant to the IF as more important than starting from a properly-matched massive halo.

The sightlines we use were drawn at a single redshift and thus do not include dynamical evolution during radiative transfer, similar to the approach of past works (e.g. \citealt{BH07}). The timescale of dynamical effects due to photoheating by the quasar can be approximated as the Jeans length of the gas $L_J$ at its initial temperature $T_\mathrm{cold}$ divided by the sound speed of the gas $c_s$ at its final temperature $T_\mathrm{hot}$. Approximating $L_J$ as $c_s H^{-1}$ for cold gas at the cosmic mean density, we have
\begin{eqnarray}
t_\mathrm{dyn}&\sim& \frac{L_J}{c_s} \sim \Delta^{-1/2} H^{-1} \left(\frac{T_\mathrm{cold}}{T_\mathrm{hot}}\right)^{1/2} \nonumber \\
&\sim& 100\ \mathrm{Myr} \times \Delta^{-1/2} \left(\frac{T_\mathrm{cold}}{500 \mathrm{K}}\right)^{1/2} \left(\frac{T_\mathrm{hot}}{3\times10^4 \mathrm{K}}\right)^{-1/2},
\end{eqnarray}
at $z=7.1$. This timescale is much longer than the ionization timescale of the gas inside the IF, so dynamical evolution will not significantly affect the progress of the IF through the IGM or the resulting IF \lya emission.

We estimate the clumping factor of the simulation by computing $\langle n_\mathrm{H}^2 \rangle / \langle n_\mathrm{H} \rangle^2$ for all the sightlines combined, masking out the region within 1 Mpc of the host halo to avoid overestimating the global clumping, and find $C\sim350$. This high clumping factor is not representative of most of the volume the IF probes in our simulations. Instead, it is dominated by rare collapsed systems with $\Delta \gg 100$. Considered individually, most of the sightlines have a clumping factor an order of magnitude smaller, closer to the $C=35$ assumed by C08. We show later in Section \ref{sect:lyaresults} that the clumping factor is not especially relevant to the IF \lya emission.

The results presented in the rest of the paper assume that the resolution of the simulation is sufficient to characterize density fluctuations on the (density-dependent) scale of the IF emission region. This is not a trivial assumption -- the presence of significant ``sub-grid" gas clumping could negate the arguments in Section \ref{sec:analyticdesc} against the use of a clumping factor. Fortunately, in Section \ref{sect:subgrid} we find that the following results are robust to the mass resolution and temperature floor of the simulation.

\subsubsection{Results for individual sightlines}\label{sect:inhomogresults}
\begin{figure}
\begin{center}
\resizebox{8cm}{!}{\includegraphics{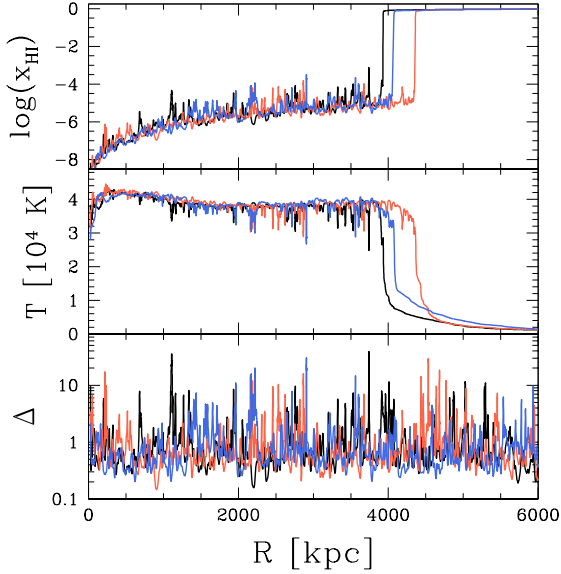}}\\
\end{center}
\caption{Output from the inhomogeneous RT model with $\alpha_\mathrm{Q}=1.5$ and $t_\mathrm{code}=25$ Myr showing neutral fraction $x_\HI$ (top panel), temperature $T$ (middle panel), and density relative to the cosmic mean $\Delta$ (bottom panel). Colors indicate the three different sightlines included in this figure, which represent typical sightlines through the IGM.}
\label{fig:rtsights}
\end{figure}

Figure \ref{fig:rtsights} shows the ionization, temperature, and density for three typical density sightlines with $\dot{N}_\mathrm{ion} = 10^{57}$ s$^{-1}$ and $\alpha_\mathrm{Q} = 1.5$ (typical of bright high-$z$ quasars) at $t_\mathrm{code} = 25$ Myr. The resulting structure is not surprising -- it largely resembles the uniform case described above. In detail, regions with higher density are somewhat cooler and have a higher equilibrium neutral fraction. The former is due to the effect of line cooling within the IF, as mentioned in Section 2.

\begin{figure}
\begin{center}
\resizebox{8cm}{!}{\includegraphics{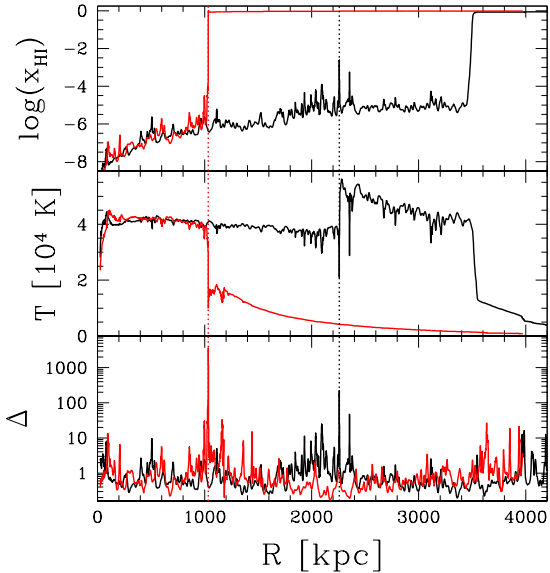}}\\
\end{center}
\caption{Output from the inhomogeneous RT model (see Figure \ref{fig:rtsights}) for atypical sightlines. The black curves show a sightline that encounters a overdensity which leaves behind a partially neutral absorber system with continuum optical depth at the hydrogen ionizing edge of $\sim1$, located at the vertical dotted black line, while the red curves show a sightline that is halted by a LLS, located at the vertical dotted red line.}
\label{fig:rtlls}
\end{figure}

A handful of sightlines encounter much higher overdensities and behave in a qualitatively different manner. The black curves in Figure \ref{fig:rtlls} show a sightline with a $\Delta\sim200$ overdensity that remains a substantial absorber of ionizing photons with optical depth at the \HIa ionizing edge $\tau_\HI\la1$ after the IF passes through it (a ``partial" Lyman limit system; pLLS), demonstrating the effect of spectral hardening on the resulting temperature. The red curves show a sightline with a $\Delta\sim3000$ overdensity that is optically thick and halts the IF (a Lyman limit system; LLS). Within $t_\mathrm{code} = 25$ Myr (corresponding to $t_\mathrm{LC}\sim50$ Myr for $\theta=\pi$), we find five sightlines each encountering pLLS and LLS, suggesting that $\la10\%$ of all quasar sightlines are affected by such systems. The frequency of pLLS and LLS increases with time as the quasar radiation field is diluted, so the majority of the atypical effects occur later than $t_\mathrm{LC}\sim40$ Myr.

\section{Analytic Method}\label{sect:analyticmethod}
The numerical method presented in the previous sections is too computationally intensive to explore the average properties of an ensemble of sightlines with varying source parameters or to construct three-dimensional maps of \lya emission properties in a simulation, motivating a more computationally efficient method to calculate the \lya emission from the IF.\footnote{Each RT sightline, run for 100 Myr without the causal correction, takes $\sim30$ minutes to compute with one 2.93 GHz Intel Xeon core on a Mac Pro compared to $\sim2$ \emph{seconds} with the analytic model described in this section.}
Moreover, the input physics is straightforward enough that one should hope for a deeper physical understanding by constructing a simplified model. In this section, we describe an ``analytic" model based on the discussion in Section 2 that reproduces the IF \lya emission from the radiative transfer model to $\sim20\%$ accuracy.

\subsection{Basic model}\label{sect:analyticbasic}
As demonstrated in Section 2, the properties of the IF can be described by a handful of simple analytic expressions. First, disregarding recombinations, the propagation of an IF into a neutral homogeneous medium can be described by balancing the number of ionizing photons and neutral atoms:
\begin{equation}
n_\HI dV = \dot{N}_\mathrm{ion} dt,
\end{equation}
which can be restated as
\begin{equation}\label{eqn:ifpropag}
\frac{dR}{dt} = \frac{\dot{N}_\mathrm{ion}}{4\pi R^2 n_\HI}.
\end{equation}
Following this expression, it is then possible to step the IF in time along a sightline with grid cells $i$ (and corresponding physical quantities $n_{\HI,i},R_i,dR_i$):
\begin{equation}\label{eqn:dti}
dt_i = \frac{4\pi R_i^2 dR_i n_{\HI,i}}{\dot{N}_\mathrm{ion}}.
\end{equation}
This method largely reproduces the IF radius as a function of time from the full numerical model, albeit with subtle differences due to secondary ionizations that will be discussed later.

The next step is to determine the \lya emission from the IF. To do this we follow the discussion in Section 2 and assume a simple physical picture for the IF where it is a structure of fixed hydrogen column density $N_\mathrm{IF}$, $x_\HI = 0.5$, and effective emission temperature $T_\mathrm{IF}$ that is a function of the IF velocity $v_\mathrm{IF}$ and density. The implied \lya surface brightness is then
\begin{equation}
SB_{\mathrm{Ly}\alpha} \sim \frac{1}{\pi} (1+z)^{-4}\langle \epsilon_\lyam \rangle_\IF dR_\IF,
\end{equation}
where $\langle \epsilon_{\mathrm{Ly}\alpha} \rangle_\IF$, the average emissivity of \lya photons, is
\begin{equation}
\langle \epsilon_{\mathrm{Ly}\alpha} \rangle_\IF \sim \frac{1}{4}\langle n_\mathrm{H}^2 \rangle_\IF E_\lyam q_\mathrm{eff}(T_\mathrm{IF}), 
\end{equation}
$\langle\rangle_\mathrm{IF}$ denotes a spatial average within the IF, and $E_\lyam=h\nu_\alpha$ is the energy of a \lya photon. The $1/\pi$ in the surface brightness equation comes from the \lya emission being directed towards the observer as it escapes the optically thick IF \citep{GW96}, and the $1/4$ in the equation for the IF emissivity comes from $x_\HI (1-x_\HI)$ assuming $x_\HI = 0.5$.

The simplest model for the temperature within the IF would be to assume a fixed value. In the absence of significant line cooling within the front, that would be a good approximation. However, that is in general not the case, and instead, to first order as in Section \ref{sec:analyticdesc}, the IF temperature can be approximated as
\begin{equation}\label{eqn:basicT}
T_\mathrm{IF} = T_\mathrm{max} - \frac{2}{3}\frac{\epsilon_\lyam}{n_\mathrm{H} k_B}t_\mathrm{IF},
\end{equation}
where $T_\mathrm{max}$ is the maximum temperature to which the gas can be heated while the IF passes through it, which depends only on the ionizing spectrum, and $t_\IF$ is the time that a parcel of gas spends inside the IF. However, due to the highly non-linear dependence of $\epsilon_\lyam$ on temperature, this approximation breaks down quickly above overdensities of a few.

This model for the \lya emission largely reproduces our radiative transfer results within a factor of a few and contains the most important physics for understanding the IF. However, it is possible to instead reproduce the radiative transfer results to within $\sim20$\% by including additional physics.

\subsection{Advanced model}\label{sect:analyticfull}
\subsubsection{Secondary ionizations}\label{sect:analytic2nd}
The propagation of the IF as a function of time is not fully described by equation (\ref{eqn:ifpropag}) above, in part due to secondary ionizations by photoelectrons. There are two dominant aspects to this. First, the effective number of ionizing photons from the source is increased due to the presence of extra ionizations close to the IF. This is evident in Figure \ref{fig:secondtemp}, where the model neglecting secondary ionizations has not propagated as far (the IF is located at the steep decline near $R\sim3$ Mpc). Second, high-energy photons ``pre-ionize" the surrounding medium, decreasing the effective neutral fraction and thus letting the IF travel faster. We have found that these two effects can be modeled by writing the effective number of ionizing photons as
\begin{equation}\label{xfactor}
\dot{N}_\mathrm{ion,eff} = \dot{N}_\mathrm{ion,0} (1 + Xt),
\end{equation}
where $\dot{N}_\mathrm{ion,0}$ is several percent larger than the actual value for the input source and $X$ is a free parameter of order a few percent per $10$ Myr for $\alpha_\mathrm{Q} = 1.5$. This simple model reproduces the size of the ionized region to nearly the spatial resolution of the input density field in most cases, once the new parameters are properly calibrated.

\subsubsection{IF temperature}\label{sect:analytictemp}
While the first order IF temperature approximation of equation (\ref{eqn:basicT}) is reasonable for small $R$ and $\Delta$, it completely breaks down when the IF slows as it passes through $\Delta \ga 10$ regions. This is because the cooling rate is a strong function of temperature (see Figure 1 of C08), so the gas within the IF will not cool indefinitely at its initial rate. Further intuition can be gained by approximating the excitation cooling rate coefficient $q_\mathrm{eff}$ as a power law in temperature over the relevant range and then writing down a simple form for the temperature evolution:
\begin{eqnarray}
\frac{dT}{dt} &\sim& -\frac{2}{3}\frac{\epsilon_\lyam}{n_\mathrm{H} k_B} \sim -\frac{1}{6}\frac{n_\mathrm{H}E_\lyam}{k_B} q_\mathrm{eff}(T) \nonumber \\
 &\sim& -\frac{1}{6}\frac{n_\mathrm{H}E_\lyam}{k_B} q_{\mathrm{eff},0} \left(\frac{T}{T_\mathrm{max}}\right)^{\alpha_T}.
\end{eqnarray}
The resulting temperature of the IF can then be approximated by analytically integrating $dT/dt$ over the time a parcel of gas spends within the IF, $t_\mathrm{IF} = dR_\mathrm{IF}/v_\mathrm{IF}$. Assuming an initial temperature $T_\mathrm{max}$, the solution is
\begin{equation}
T_\mathrm{IF} \approx T_\mathrm{max} \left(\frac{T_\mathrm{max}}{(\alpha_T-1)E_\lyam q_\mathrm{eff,0}n_\mathrm{H} t_\mathrm{IF}/6k_B+T_\mathrm{max}}\right)^{\frac{1}{\alpha_T-1}}.
\end{equation}
For our fiducial set of simulations with $\alpha_\mathrm{Q}=1.5$, we find $T_\mathrm{max}\sim 3.4\times10^4$ K, $q_\mathrm{eff,0}\sim4\times10^{-10}$ cm$^3$ s$^{-1}$, and $\alpha_T\sim8.5$. This best-fit power law cooling rate function is very similar in character to the input cooling rate in the radiative transfer model, suggesting that this simplified approach to cooling within the IF is a reasonable one.

\subsubsection{Recombinations}\label{sect:recombs}
The model described above does not account for loss of ionizing photons to recombinations within the ionized region. In most cases this will not be important because $t_\mathrm{rec} \gg t_\mathrm{LC}$. However, sufficiently dense regions ($\Delta \ga 100$) can remain substantially neutral and have residual optical depths at the hydrogen ionizing edge of order unity or higher (see Section \ref{sect:inhomogresults}). The IF beyond these regions will then proceed more slowly. We include this effect by reducing the number of ionizing photons available to expand the ionized region by the number of recombinations along the sightline,
\begin{equation}
\dot{N}_\mathrm{rec} = \int_0^{R_\IF} \alpha_\HII^A n_\mathrm{H}^2 4\pi r^2 dr,
\end{equation}
where $\alpha_\HII^A$ is the case A recombination coefficient and we assume the ionized region has $x_\HII \approx 1$. The temperature of the ionized gas is assumed to be $T_\mathrm{final} = T_\mathrm{IF} + T_\HeIII$ where $T_\HeIII = 7000\times(1.5/\alpha_\mathrm{Q})$ K is an approximation of the additional heating due to the second ionization of helium based on the results of the radiative transfer model. This approximation for the IGM temperature is very rough, however, as the gas in the full simulation does not remain at a fixed temperature but instead cools due to adiabatic and inverse Compton cooling. Fortunately, the recombination rate is not an especially strong function of temperature, so the approximation is not a particularly bad one. Some of the systematic error in the recombination rate correction to the propagation of the IF is corrected by calibrating the $X$ factor described in Section \ref{sect:analytic2nd}.

\subsection{Calibrating fit parameters}\label{sect:analyticparams}
The final analytic model has six free parameters, \{$N_\IF,\dot{N}_\mathrm{ion,0},X,q_\mathrm{eff,0},T_\mathrm{max},\alpha_T$\}, and we find the best fit set of parameters by minimizing the squared residuals of $SB_\mathrm{Ly\alpha}(t)$ compared to the radiative transfer model over a set of six sightlines. In Table \ref{tab:amparams} we list the best fit parameters for radiative transfer models with varying quasar spectrum, where the $\alpha_\mathrm{Q} = 1.5$ model is our fiducial one.

\begin{table}
\begin{center}
\caption{List of best-fit analytic model parameters for different quasar spectrum power law indices $\alpha_\mathrm{Q}$.}
\label{tab:amparams}
\begin{tabular}{c c c c c c c}
\hline \noalign {\smallskip}
$\alpha_\mathrm{Q}$ & $N_\IF$$^a$ & $\dot{N}_\mathrm{ion,0}/\dot{N}_\mathrm{input}$ & $X$ & $T_\mathrm{max}$$^b$ & $q_\mathrm{eff,0}$$^c$ & $\alpha_T$ \\
\hline \noalign {\smallskip}
1.3 & 9.04 & 1.15 & 0.02 & 4.0 & 7.6 & 7.5 \\
1.4 & 6.23 & 1.10 & 0.018 & 3.7 & 5.6 & 7.75 \\
1.5 & 4.83 & 1.07 & 0.015 & 3.4 & 4.1 & 8.5 \\
1.6 & 3.74 & 1.035 & 0.01 & 3.2 & 3.0 & 9.0 \\
1.7 & 3.12 & 1.02 & 0.005 & 3.0 & 2.8 & 10.0 \\
1.8 & 2.71 & 0.99 & 0.002 & 2.85 & 2.0 & 11.0 \\
\hline \noalign {\smallskip}
\end{tabular}
\end{center}
$^a$ $10^{18}$ cm$^{-2}$\\
$^b$ $10^4$ K\\
$^c$ $10^{-10}$ cm$^3$ s$^{-1}$
\end{table}

\subsection{Comparison to radiative transfer results}\label{sect:analyticvsrt}
\begin{figure}
\begin{center}
\resizebox{8cm}{!}{\includegraphics{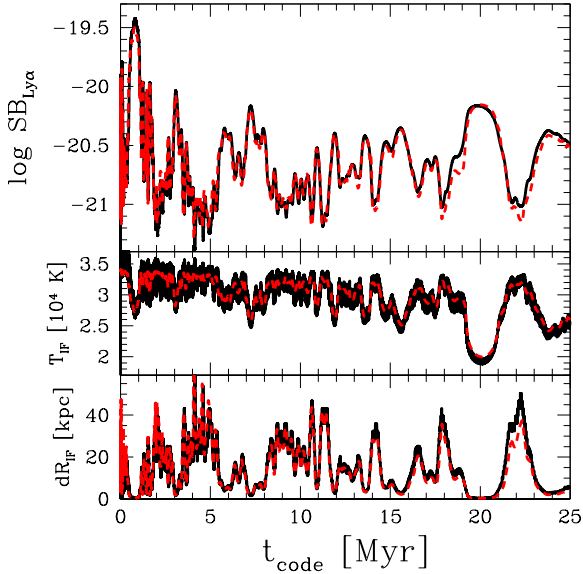}}\\
\end{center}
\caption{Comparison between the radiative transfer model (solid black) and analytic model (dashed red) with no causal correction applied. Upper panel: plane-parallel \lya surface brightness. Middle panel: temperature within the IF. Bottom panel: width of the IF.}
\label{fig:amcompare}
\end{figure}

Figure \ref{fig:amcompare} compares the analytic model to the radiative transfer model for a representative sightline. $T_\IF$ is the temperature of the gas in the center of the IF, where $x_\HI\sim0.5$, while $dR_\IF$ is its width. In the radiative transfer simulations, we define $T_\IF$ as the temperature of the first grid cell from the origin with $x_\HI>0.5$ and $dR_\IF$ as the distance between cells with $x_\HI=0.05$ and $x_\HI=0.75$. These simple definitions were chosen to act as rough approximations to the general properties of the IF and they are not meant to be exact. The analytic model simultaneously matches the surface brightness, IF temperature, and IF width very well. However, some slight deviations remain, such as the dip in $SB_\lyam$ at $t_\mathrm{code}\sim19$ Myr. This is due to the pre-ionization region (Section \ref{sect:secondion}) extending into relatively high density gas, which causes a small excess of \lya emission relative to the analytic model.

\begin{figure*}
\begin{center}
\resizebox{8cm}{!}{\includegraphics{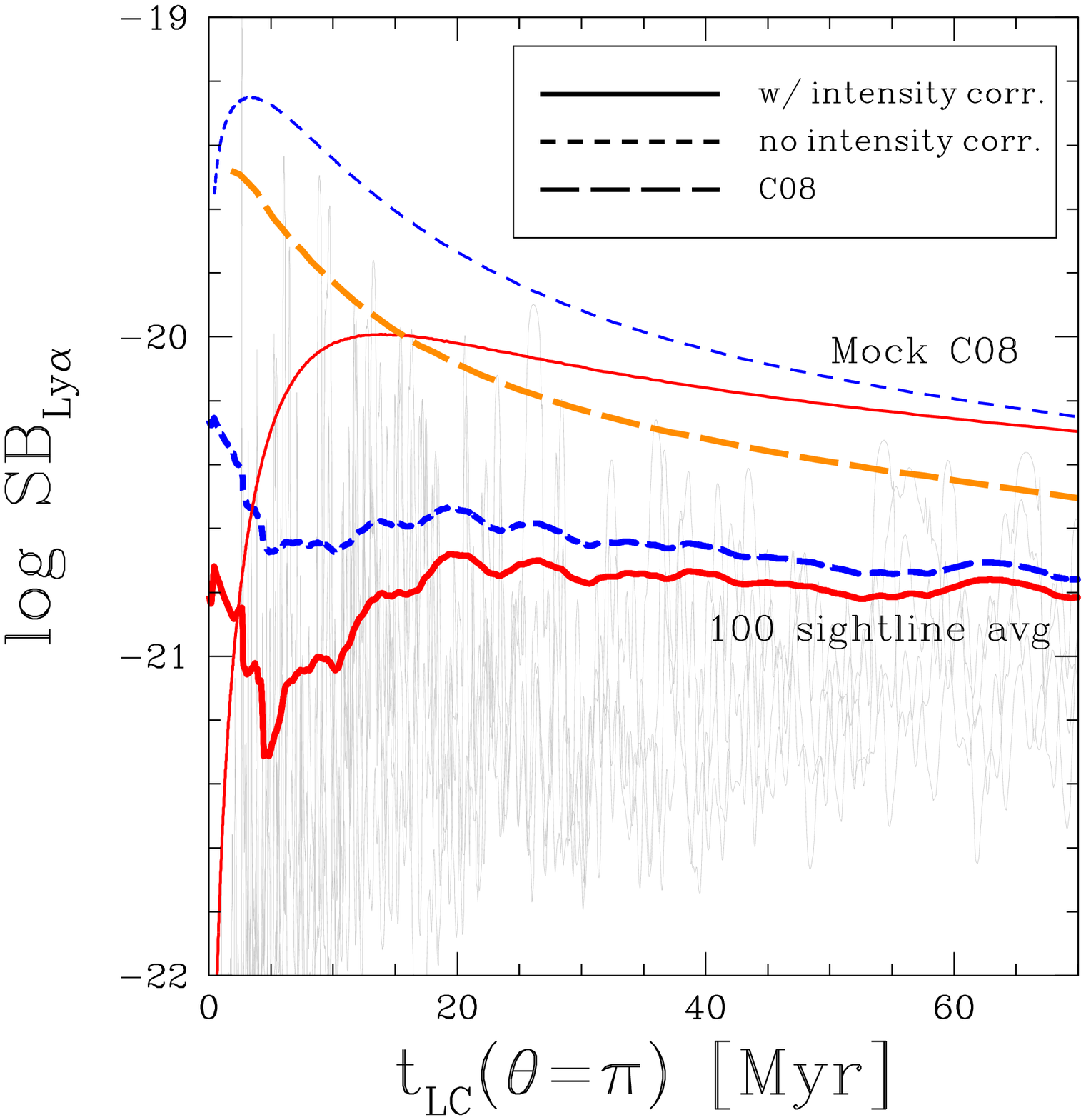}}
\hspace{0.13cm}
\resizebox{8cm}{!}{\includegraphics{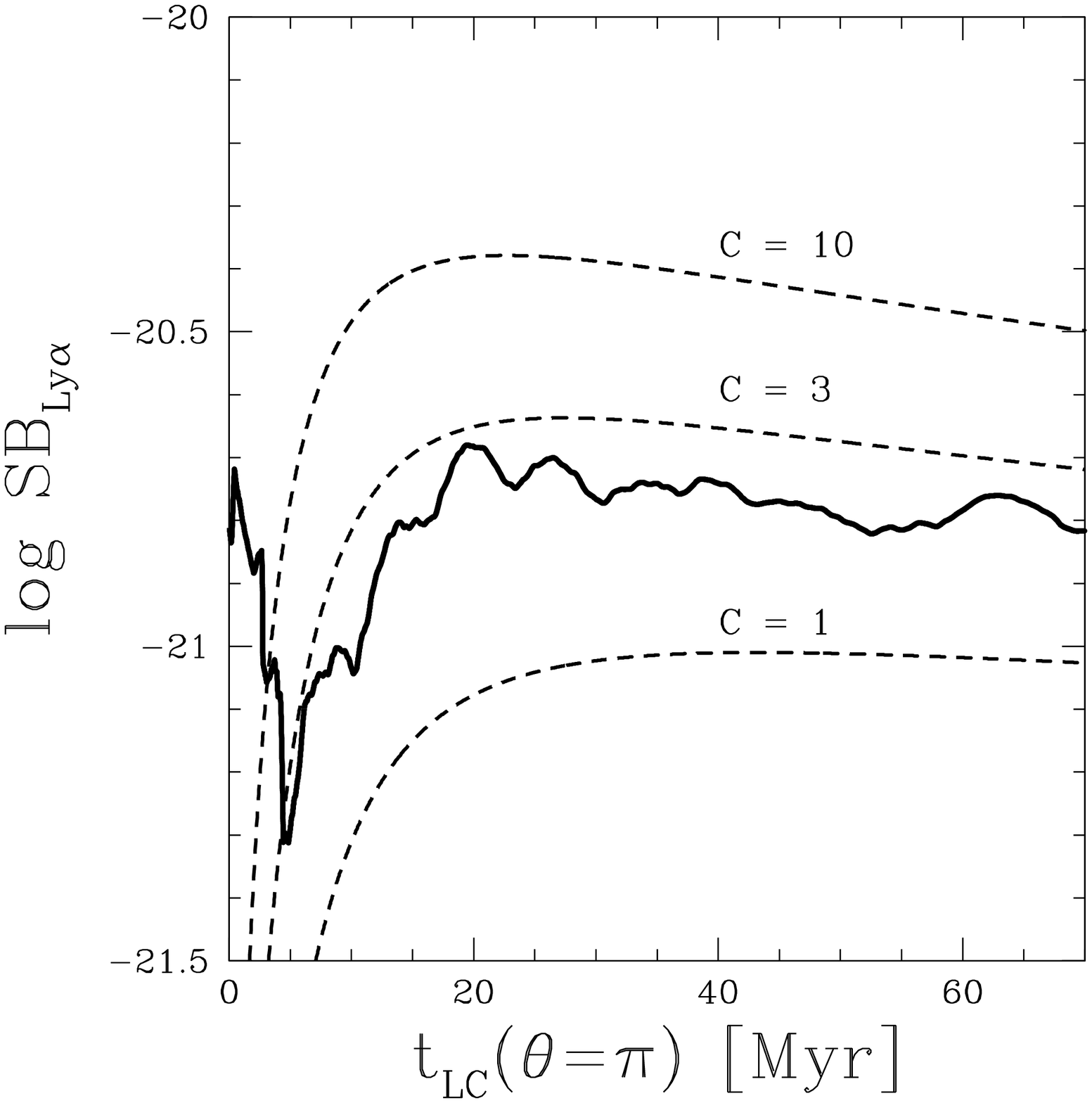}}
\end{center}
\caption{\emph{Left}: Plane-parallel \lya surface brightness for 10 individual sightlines (thin gray curves) and a 4 Myr top-hat smoothed average of all 100 sightlines (thick curves) with or without the causal intensity correction (solid and short-dashed, respectively) compared to the model prediction from C08 scaled to $z=7.1$ (long-dashed curve) and our calculation using the C08 input parameters (upper thin curves). \emph{Right}: Dashed curves show the predicted \lya surface brightness for the uniform model calculation assuming different values of the clumping factor ($C=1,3,10$ from bottom to top) compared to the inhomogeneous IGM average (solid curve). Note the different scale on the vertical axis from the left panel.}
\label{fig:ppresults}
\end{figure*}

The analytic model was calibrated to $t_\mathrm{code}=25$ Myr output from the radiative transfer model (corresponding to $t_\mathrm{LC}\sim40$ Myr), and thus it does not match as well at later times. Specifically, by comparing to a 100 Myr radiative transfer simulation, equation (\ref{xfactor}) becomes a poor approximation to the effective number of ionizing photons after $t_\mathrm{code} \sim40$ Myr. Similarly, the recombination rate assumes a constant IGM temperature, but in the radiative transfer model the gas continues to cool after the ionization of \HeIIa due to adiabatic and Compton cooling, slowly increasing the true recombination rate relative to the approximation in Section \ref{sect:recombs}.

Another limitation involves the effect of secondary ionizations on the IF with a hard source spectrum. For relatively hard source spectra, the pre-ionization region ahead of the IF becomes an important source of \lya photons. Additionally, the assumption of a constant density within the IF becomes less valid because the wider IF will encompass a range of densities. Finally, the analytic model assumes a single power law for the effective cooling rate as a function of temperature, which is not a good assumption at the higher temperatures inside the IF of a hard spectrum source. We find that the analytic model no longer provides a reasonable approximation to the surface brightness for $\alpha_\mathrm{Q} \la 1.3$.

The analytic model ignores the gradual hardening of the spectrum reaching the IF due to absorption of ionizing photons by residual neutral atoms within the ionized region. This hardening broadens the IF slightly over time and change the balance of heating and cooling, causing the analytic model to drift from the radiative transfer model increasingly over time. However, this deviation is only at the few percent level after $t_\mathrm{code}\sim50$ Myr ($t_\mathrm{LC}\sim90$ Myr), so it may be unimportant if the extremely luminous quasar phase is limited to a few tens of Myr (e.g. \citealt{TS13}).

We find that the analytic model is a reasonable approximation to the RT model for $\alpha_\mathrm{Q} \ga 1.3$ and $t_\mathrm{LC} \la 75$ Myr, and only for the ``typical" 90\% of sightlines that do not encounter optically thick absorbers.

\subsection{Causal correction in the analytic model}
After calibrating the analytic model to reproduce the infinite speed of light radiative transfer results, we applied the causal correction described in Section \ref{sect:causal} by transforming each $dt_i$ in equation (\ref{eqn:dti}) to its counterpart $dt_{\mathrm{LC},i}$ observed on the light cone with 
\begin{equation}
dt_{\mathrm{LC},i} = dt_i (1+\frac{dR/dt}{c}(1-\cos{\theta})),
\end{equation}
as in equation (\ref{eqn:dtconversion}), and correcting the output surface brightness following equation (\ref{eqn:sbconversion}),
\begin{equation}
SB_\lyam \rightarrow SB_\lyam \left(1 + \frac{dR/dt}{c}(1-\cos{\theta})\right)^{-1}.
\end{equation}

As discussed in Section \ref{sect:causal}, for our main results we do not apply the causal correction directly to the output of the radiative transfer model due to its much more stringent requirements on spatial and temporal resolution. We have confirmed with small scale tests that the above corrections to the time steps and surface brightness are equivalent to applying the coordinate transformation in equation (\ref{eqn:caustime}) directly to the radiative transfer model, albeit at much greater computational expense.

\section{\lya Surface Brightness}\label{sect:lyaresults}
An accurate determination of the observable properties of the IF \lya emission will depend sensitively on the detailed \lya radiative transfer both out of the IF and within the ionized region close to the quasar. However, these processes only serve to disperse \lya photons out of the line of sight, so a robust upper limit on the \lya surface brightness can be found by assuming a plane-parallel geometry. That is, assuming that the IF behind the quasar is an optically thick slab, the emergent intensity is (\citealt{GW96}, C08)
\begin{equation}
I_\lyam \approx \frac{1}{\pi}\int_\IF \epsilon_\lyam dR,
\end{equation}
corresponding to an observed surface brightness $SB_\lyam = I_\lyam (1+z)^{-4}$. The overall redshift dependence of the surface brightness will be substantially weaker than $(1+z)^{-4}$ because of the dependence of $I_\lyam$ on physical density and temperature inside the IF. Ignoring the temperature dependence of the emission, we have $\epsilon_\lyam \propto n_\mathrm{H}^2$ and $dR_\IF \propto n_\mathrm{H}^{-1}$, so $I_\lyam \propto (1+z)^{3}$, leading to a surface brightness that scales as $\propto (1+z)^{-1}$. In detail, the redshift dependence depends on the evolving structure of the density field and the physical density dependence of the cooling that strongly regulates the IF emission and is not trivial to determine. In the case of a uniform medium with clumping factor $C=35$, C08 found that $SB_\lyam \propto (1+z)^{-2}$, and we find a similar scaling applies to our calculations as well.

There are several effects that this simple plane-parallel picture ignores. We have assumed that, because of the neutral hydrogen gradient within the IF, \lya photons will preferentially escape towards the source with zero photons escaping in the IF propagation direction \citep{GW96}. The emission may be emitted more isotropically when the IF is moving at relativistic speeds and scattered by residual neutral hydrogen in the ionized region, leading to suppression by up to a factor of four (C08). Also, the effective area of a given IF shell segment will depend on the angle from the line of sight through projection of the emitting surface, which is non-trivial in the clumpy IGM. Finally, \lya photons emitted from the IF may be scattered by the foreground neutral IGM if they have not had a long enough path length to redshift out of line center \citep{ME98}. For a photon emitted from the IF behind the quasar, the path length will be given by the sum of the causally corrected IF radius and the uncorrected IF radius. We find that the IGM damping wing absorption could suppress the \lya emission by $\sim15$--$35$\% for $5 < t_\mathrm{LC} < 30$ Myr, significantly larger than suggested by C08\footnote{It appears \citet{C08} ``double-counted" the light-travel distance from the far side of the ionized region to the near side; the causality-corrected shape of the ionized region already reflects the light-travel distance by definition \citep{Yu05}.} but still not especially severe.

We ignore these effects in the following discussion and focus on the plane-parallel estimate of the surface brightness which is almost certainly an \emph{overestimate} of the true emission. In the next section we will find that even this optimistic overestimate is significantly fainter than previously predicted and is likely inaccessible to current telescopes, making further modeling less interesting.

\subsection{Results}
In the left panel of Figure \ref{fig:ppresults} we compare our plane-parallel surface brightness calculation to the similar model from C08 for \lya emission from behind the quasar ($\theta=\pi$). The rapidly varying thin curves show the effective $SB_\mathrm{Ly\alpha}$ as a function of time for ten of the 100 sightlines. However, those $SB_\mathrm{Ly\alpha}(t)$ are not representative of the emission averaged on large scales, as one would hope to measure. To approximate an observation that would average over the density field structure in 3D, the thick curves in Figure \ref{fig:ppresults} show a 4 Myr top-hat smoothed average of all 100 sightlines, with (solid) and without (dashed) the causal correction to the IF \lya intensity from equation (\ref{eqn:sbconversion}). Our resulting \lya surface brightness is substantially lower than the C08 model (long-dashed curve) scaled to $z=7.1$ by $SB_\lyam\propto(1+z)^{-2}$ (C08). The short-dashed curves in the right panel of Figure \ref{fig:ppresults} show uniform density models with varying clumping factor. By comparing the uniform density models to the inhomogeneous model, we find the ``effective" clumping factor of the inhomogeneous IGM, as probed by the \lya emission, to be $C_\mathrm{eff} \sim 2.5$, compared to $C\sim350$ for the density field. The enhancement in the \lya emission at $t\la5$ Myr is due to the local overdensity of the host halo and has been broadened by the smoothing. The timescale over which the emission is enhanced by the host halo is $\sim2$ Myr, corresponding to a local overdensity scale of $\sim300$ kpc.

The low effective clumping factor relative to the density field is the result of a combination of several effects acting to increase and decrease the $n_\mathrm{H}^2$ boost from a clumpy IGM. As discussed previously, the thinness of the IF relative to the scale of density fluctuations causes the \lya emission to vary proportionally to $n_\mathrm{H}$ instead of $n_\mathrm{H}^2$. The extra cooling in dense regions further weakens the dependence on $n_\mathrm{H}$. However, ignoring causal effects, the IF will typically spend more time inside of dense regions because they take longer to ionize. On the other hand, the causal correction reduces the IF velocity more strongly within underdense regions, which increases their contribution relative to dense regions where $v_\IF$ is smaller relative to $c$.

We have also computed $SB_\lyam$ using the same input parameters as C08 ($\alpha_\mathrm{Q}=1.7$, $C=35$, no secondary ionizations) to directly compare the differences between our methods, shown by the thin curves at the top of Figure \ref{fig:ppresults} where the solid and dashed curves are with and without the causal correction, respectively. The dashed curve is directly comparable to the long-dashed curve from C08, while the solid curve shows the additional effect of the causal correction on $SB_\lyam$ that we have highlighted. Interestingly, our (intensity uncorrected) model predicts \lya emission enhanced by a constant factor of $\sim2$ relative to C08. C08 did not include soft X-ray radiation in their simulations, which may cause their IF to be narrower and reduce the heat input into the IGM, especially in the regions beyond the IF (compare Figure 5 of C08 to Figure \ref{fig:unifall} in this work). Because the collisional excitation rate is such a strong function of temperature (see Figure 1 of C08), only a modest $\sim10\%$ difference in IF temperature is enough to fully account for the discrepancy.

\begin{figure}
\begin{center}
\resizebox{8cm}{!}{\includegraphics{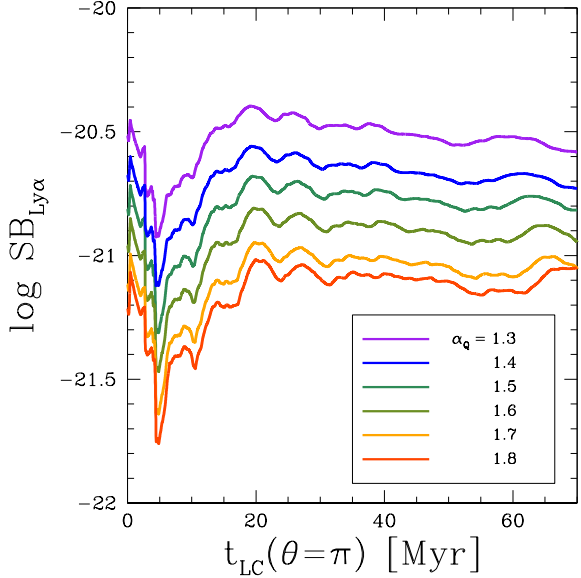}}\\
\end{center}
\caption{Plane-parallel \lya surface brightness for a 4 Myr top-hat smoothed average of all 100 sightlines for a range of different quasar spectra with power law indices of $\alpha_\mathrm{Q} = 1.3,1.4,1.5,1.6,1.7,1.8$ from top to bottom with fixed $\dot{N}_\mathrm{ion}$.}
\label{fig:alphasb}
\end{figure}

In Figure \ref{fig:alphasb}, we compare the predicted plane-parallel surface brightness for a range of input quasar spectra with varying power law index and fixed $\dot{N}_\mathrm{ion}$. Like C08, we find that the surface brightness is sensitive to the quasar ionizing spectrum. This sensitivity is due to a combination of changes in the IF width and the IF temperature, which act together to increase or decrease the \lya emission. While the analytic model is not applicable to quasar spectra with $\alpha_\mathrm{Q} \la 1.3$, we can still estimate the effect of a harder spectrum with the radiative transfer code. We find that for $\alpha_\mathrm{Q}=1.0$ $(0.5)$ the total \lya emission is roughly 6 (20) times brighter than our fiducial $\alpha_\mathrm{Q}=1.5$ case.

\begin{figure}
\begin{center}
\resizebox{8cm}{!}{\includegraphics{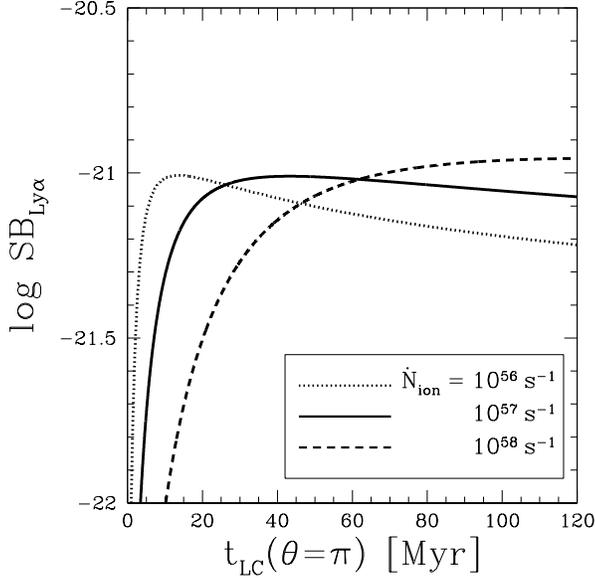}}\\
\end{center}
\caption{\lya surface brightness in the uniform model (with $C=1$) for varying quasar ionizing luminosities. The higher luminosity source is brighter at late times, but suffers from an enhanced causal correction at early times.}
\label{fig:nionsb}
\end{figure}

By computing the \lya emission from cosmological simulation sightlines at higher redshifts, we find that, in agreement with C08, the surface brightness scales as $(1+z)^{-2}$. However, due to the causal correction factor dependence on IF velocity, the relationship between surface brightness and $\dot{N}_\mathrm{ion}$ is no longer one-to-one. Without this velocity dependent correction, the dominant effect of $\Nion$ is on the cooling within the IF: larger $\Nion$ means less time spent within the IF which means less cooling, and vice versa. In our model this effect is largely neutralized by the causal correction factor. When $\Nion$ is larger the IF is somewhat hotter and thus more luminous, but it is also moving faster relative to $c$ so the \lya intensity is diminished. Figure \ref{fig:nionsb} demonstrates this effect over the range $10^{56}$ s$^{-1}<\Nion<10^{58}$ s$^{-1}$. Note, however, that while a fainter source may be somewhat brighter at quasar-relevant timescales ($t_\mathrm{LC} \sim 10$ Myr), the emission will cover a smaller region on the sky and experience enhanced IGM damping wing absorption due to the smaller size of its ionized region, so they may not provide a more promising target for observations.

\begin{figure}
\begin{center}
\resizebox{8cm}{!}{\includegraphics{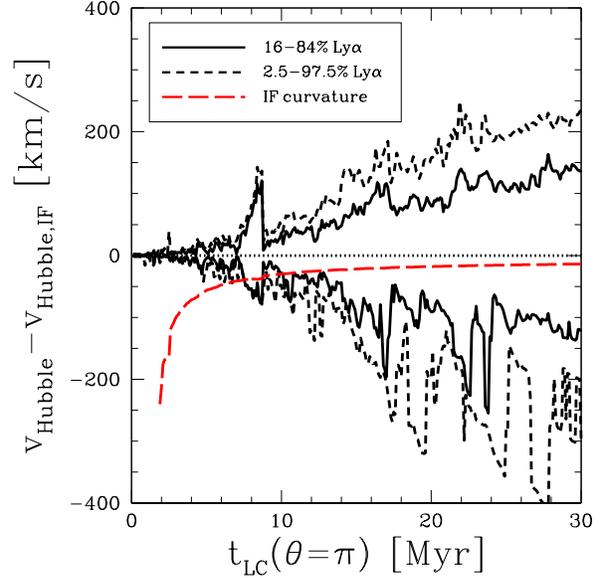}}\\
\end{center}
\caption{Velocity width of IF \lya emission due to the spread in Hubble flow velocities of the IF. Solid and short-dashed curves show the width of the central 16-84\% and 2.5-97.5\% of the total \lya emission, respectively, neglecting the intrinsic line width. The long-dashed curve shows the velocity shift due to curvature of the IF at 50 arcseconds away from the quasar on the sky. The zero-point is set by the Hubble flow velocity of the IF.}
\label{fig:vwidth}
\end{figure}

While our model does not include \lya radiative transfer and thus cannot make specific predictions about the velocity structure of the emission, one possible contributor is the distribution of Hubble flow velocities across the IF \lya-emitting surface. That is, the ``lumpiness" of the IF due to inhomogeneities influencing the IF velocity could broaden the IF emission in velocity space. The solid and short-dashed curves in Figure \ref{fig:vwidth} show the 16-84\% and 2.5-97.5\% widths of the \lya emission profile in velocity space due to the Hubble flow relative to the \lya brightness-weighted mean velocity, ignoring the intrinsic line width and without any temporal smoothing. By $t_\mathrm{LC}\sim10$--$20$ Myr, the lumpy nature of the IF can broaden the emission profile by $\ga 100$~km/s, similar to the expected intrinsic line width (C08). Observations of a small field behind the quasar would likely see a smaller velocity width due to correlations in the density field.

Another contributor to the line width is the shift in central velocity due to curvature of the IF across the field of view of a hypothetical observation. Because the IF \lya emission is quite weak, detection would likely require integration over $\ga1$ arcmin$^2$. The long-dashed curve in Figure \ref{fig:vwidth} shows the Hubble flow velocity difference between the IF directly behind the quasar and the IF 50 arcseconds away, approximating the shape of the ionized region as spherical.\footnote{Observation of the IF along the light cone flattens the apparent shape of the ionized region behind the source \citep{Yu05}, so a spherical approximation mildly overestimates the velocity shift.} At $z\sim7$, 50 arcseconds corresponds to $\sim0.25$ Mpc. At $t_\mathrm{LC}\sim10$ Myr, the radius of the ionized region along the light cone behind the source is typically $\sim1.5$ Mpc, so this represents a small but not negligible fraction of the IF surface. As demonstrated in Figure \ref{fig:vwidth}, the curvature of the IF is likely to be unimportant compared to the lumpy structure except at early times ($t_\mathrm{LC}\la10$ Myr) or for observations covering a larger field of view.

\subsection{Is sub-grid clumping important?}\label{sect:subgrid}
The analysis above assumes that our cosmological simulation provides an accurate representation of neutral gas clumping in the IGM on the scale of the IF. With a gas particle mass of $\sim10^6$ M$_\odot$, our simulation should resolve the Jeans mass of 500 K neutral gas at the mean density, $M_J\sim 5\times10^7$ M$_\odot$. Although 500 K is a very reasonable temperature estimate for the IGM at $z\sim7$ and is below several models (e.g., \citealt{FurlanettoXray,PL10}), the lack of constraints on IGM heating at very high redshifts may mean that it is an overestimate, which would allow for clumping on even smaller scales. If substantial clumping existed on scales much smaller than the IF, the arguments in Section \ref{sec:analyticdesc} against the use of the clumping factor would no longer hold, and the expected \lya emission could increase.

To investigate the potential effect of enhanced clumping on our results, we ran an additional cosmological simulation (with the same cosmology) in a volume 3 Mpc$/h$ on a side with a minimum temperature of 50 K, compared to 12.5 Mpc$/h$ and 500 K, respectively, for the fiducial simulation. We then drew 50 randomly-placed sightlines through the density field and performed the same analysis as described in the previous section to calculate the expected \lya surface brightness. Despite the significantly enhanced clumping compared to the fiducial simulation, the time-averaged \lya surface brightness is nearly identical to Figure \ref{fig:ppresults}.

While the IF in the second simulation is larger than the Jeans length of $\Delta=1$ gas, most of the time-averaged \lya emission comes from modestly overdense regions with $\Delta\ga5$. Because the IF width scales more strongly with density than the Jeans length, $dR_\IF \propto \Delta^{-1}$ versus $L_J \propto \Delta^{-1/2}$, it is relatively easy for the IF to resolve density fluctuations in overdense regions. The Jeans length of 50 K (500 K) neutral gas at $z\sim7$ is roughly equal to the IF width for $\Delta\sim6$ (2), so fluctuations in the density field above this threshold will be resolved by the IF. More quantitatively, one can estimate the relevant sub-IF-scale clumping factor,
\begin{equation}
C_\IF = \frac{\frac{1}{dR_\IF}\int_\IF n_\mathrm{H}^2 dR}{\left(\frac{1}{dR_\IF}\int_\IF n_\mathrm{H} dR\right)^2},
\end{equation}
along a sightline through the density field, where the integration over the IF is defined by a region with column density $N_\IF$ as in the analytic model of Section \ref{sect:analyticmethod}. Averaged over the sightlines, we find $C_\IF\sim1.1$ and $\sim1.4$ for the fiducial simulation and 50 K simulation, respectively, showing that clumping below the IF scale is unimportant even in an IGM significantly colder than expected. The highest $C_\IF$ values in the 50 K simulation are found at the edges of dense regions where there is a strong gradient in the density field across the IF width. $C_\IF\approx1$ inside the dense regions that contribute most of the \lya emission, so the effect of the spatially averaged value on the total emission is small.

\subsection{Discussion}
The \lya emission predicted by our model, even in the absence of \lya radiative transfer, is a factor of 3--15 weaker than that predicted by C08 for quasar ages $\la30$ Myr. This difference comes about due to two effects that were not considered in C08. First, the global clumping factor of a simulation does not describe the enhancement of \lya emission from density fluctuations, as described in Section \ref{sec:analyticdesc}. Second, while they included a causal correction factor to propagate the IF at the correct speed, they did not consider the effect of this causal correction on the intensity of the \lya emission (Section \ref{sect:causal}). As discussed in \citet{MR94}, relativistic IFs are narrower than their non-relativistic counterparts, with a width of roughly the speed of light times the ionization timescale $t_\mathrm{ion}$~$\sim$~$\Gamma^{-1}$ instead of the typical optical depth criterion. This further decreases the emission at early times.

Our less optimistic prediction for IF \lya emission calls its detectability with current and future instruments into question. The results in the previous section ignored \lya radiative transfer effects which decrease the observable emission by scattering \lya photons out of the line of sight. C08 found that these effects amounted to a factor of 0.58 correction to the plane-parallel calculation, predominantly due to scattering by residual neutral hydrogen inside the ionized region. They also found that the relativistic motion of the IF could reduce the emission by another factor of $\sim2$. Thus, our simplified calculation likely overestimates the \lya surface brightness by a factor of $\sim2$--$4$. Keeping this limitation in mind, our fiducial value for the observable \lya surface brightness is $10^{-21}$ erg s$^{-1}$ cm$^{-2}$ arcsec$^{-2}$. The expected signal-to-noise ratio (S/N) for an observation of a $z=7.1$ quasar field is
\begin{eqnarray}\label{eqn:signal}
\mathrm{S/N} &\sim& 0.35\,C_f SB_{-21} \left(\frac{D}{10\,\,\mathrm{m}}\right) \left(\frac{\zeta}{1.0}\right) \left(\frac{f}{0.25}\right)^{\frac{1}{2}} \left(\frac{\Delta\lambda}{3\,\,\mathrm{\AA}}\right)^{-\frac{1}{2}}\nonumber \\
&& \times  \left(\frac{t_\mathrm{obs}}{40\,\,\mathrm{hr}}\right)^{\frac{1}{2}} \left(\frac{\Delta\Omega}{600\,\,\mathrm{arcsec}^2}\right)^{\frac{1}{2}},
\end{eqnarray}
where $C_f$ is the covering factor of neutral IGM within the field of view, $SB_{-21}$ is the \lya surface brightness in units of $10^{-21}$ erg s$^{-1}$ cm$^{-2}$ arcsec$^{-2}$, $D$ is the diameter of the telescope, $\zeta$ is the atmospheric transmission, $f$ is the system efficiency, $\Delta\lambda$ is the spectral bin ($3\,\,\mathrm{\AA}\sim 100$ km/s), $t_\mathrm{obs}$ is the integration time, and $\Delta\Omega$ is the observed area of the sky. The various parameters have been chosen to mimic an observation with the upcoming Keck Cosmic Web Imager (KCWI; \citealt{Martin10}) assuming perfect sky subtraction and noise dominated by shot noise of the typical sky background between sky lines on Mauna Kea\footnote{\url{http://www.gemini.edu/sciops/telescopes-and-sites/observing-condition-constraints/ir-background-spectra}}. This low S/N is in spite of the cold, clumpy IGM from the hydrodynamic simulations which should overestimate the emission. Similar observations with a future ground-based large aperture instrument such as the Thirty Meter Telescope (TMT) could only achieve $\mathrm{S/N}\sim1$.

An alternative observing approach would be to attempt a narrowband photometric detection over a wider area. The potential scale of the \lya emission on the sky is a few arcmin$^2$, so the relevant $\Delta\Omega$ can in principle be much larger than the field of view of KCWI or similar spectrographic instruments. In that case, for $\Delta\lambda = 75$ \AA, $f = 0.5$, and $\Delta\Omega = 4$ arcmin$^2$ in equation (\ref{eqn:signal}), we find S/N $\sim 0.5$ $C_f$, which is only a marginal improvement if $C_f$ is comparable between the two cases.

The sky background at 1 $\mu$m from the ground is dominated by atmospheric emission (e.g. \citealt{SS12}), so improved S/N could potentially be achieved by observing from space. Above the Earth's atmosphere, the background is dominated by zodiacal light which can be minimized by observing a target at high ecliptic latitude. For example, the zodiacal light at 1 $\mu$m near the north ecliptic pole is roughly a factor of 20 smaller than the sky background assumed in equation (\ref{eqn:signal}) \citep{Giavalisco02}. This means that a space-based observation could potentially gain a factor of $\sim4.5$ in S/N, though almost certainly at the expense of telescope diameter.

If quasar IF \lya emission were detected by future instruments, the brightness would provide a joint constraint on the temperature of the IF (i.e. the shape of the quasar ionizing spectrum) and neutral fraction of the IGM. Additionally, the redshift of the \lya line would indicate the current size of the quasar ionized region, placing a constraint on the total number of ionizing photons emitted by the quasar. Maps of this emission would allow study of the tomography of reionization -- regions that were initially neutral will light up, while previously ionized regions will remain dark (C08). However, given the low expected S/N, this seems unlikely to be achieved in the near future. Based on our calculations, a detection of this faint \lya emission would require a high degree of IGM clumping on scales smaller than the IF, a hard quasar ionizing spectrum, or some combination of the two. Our investigation of enhanced clumping with a simulation of a colder IGM in the previous section suggests that the former is unlikely. As for the latter, for a S/N $\ga5$ detection with a thirty-meter-class telescope in 40 hours the target quasar must have $\alpha_\mathrm{Q}\la1$, ignoring any additional \lya radiative transfer effects due to the much wider IFs.

Five of the 100 sightlines we considered encountered optically thick absorbers, LLS, within $t_\mathrm{LC} \la 40$ Myr. Once the IF reaches the LLS overdensity ($\Delta \ga 1000$), the \lya surface brightness plateaus as the IF becomes a stationary ionized ``skin" on the surface. The \lya surface brightness varies with LLS distance from the quasar as $R_\mathrm{LLS}^{-2}$, as one would expect for the ``reflection" of a fixed fraction of the incoming ionizing radiation towards the observer \citep{GW96}. While the surface brightness of the LLSs is much higher than expected for the IF as a whole ($SB_\mathrm{LLS}\sim10^{-19} [R_\mathrm{LLS}/1.4$ Mpc$]^{-2}$ erg s$^{-1}$ cm$^{-2}$ arcsec$^{-2}$), the physical size of the associated density peak is only a few kpc across corresponding to $\la 1$ arcsec$^{-2}$. This could provide a modest boost to the integrated IF emission, but as static features they will be separated in velocity space from the rest of the IF fairly quickly. We stress that this is unlikely to be a very accurate description of the emission from optically thick systems because we do not consider the hydrodynamical effects of photoionization heating (e.g. evaporation). These systems are analogous to fluorescent \lya emission at lower redshift when the universe is fully ionized (e.g. \citealt{Cantalupo12,TS13}), and thus do not seem to be a compelling probe of the reionization process. However, detection of such systems may allow an independent constraint on the number of optically thick absorbers at high redshift and thus the mean free path of ionizing photons.

Due to the computational speed of the analytic model, one may envision a method to generate maps of the IF \lya emission by drawing rays through a three-dimensional cosmological density field. Such a technique would undoubtedly be faster than performing the relativistic ionizing radiative transfer that would otherwise be necessary. We neglected to pursue this further because the expected signal is too weak.

\section{Conclusion}
In this work we have explored the physics of \lya emission from quasar ionization fronts in an inhomogeneous IGM. Our radiative transfer modeling includes much of the relevant physics and resolves the heating, cooling, and ionization structure of the IF as it passes through a varying density field. In an improvement over past work, we include the effect of secondary ionizations by high-energy photoelectrons and properly account for causality in our infinite speed-of-light simulations. We have developed a relatively simple analytic framework that allows for rapid computation of \lya emission along a sightline through the IGM, and we applied this model to a ensemble of 100 sightlines through a hydrodynamic simulation. The resulting average surface brightness is a factor of 3--15 fainter than the prediction by C08, reflecting both the smaller effective clumping factor of the \lya emission and causal suppression due to the near-relativistic expansion at quasar relevant timescales ($5 \la t_\mathrm{LC} \la 30$ Myr).

Our analysis assumes that the light from the quasar reaches a completely neutral region of the IGM within its lifetime. This may be unlikely during reionization because the presence of a luminous quasar indicates a biased region of the Universe which will likely be ionized early \citep{AA07,Lidz07}. In a more realistic scenario where the quasar light does not reach neutral gas for $\ga5$ Mpc, the \lya emission would not ``switch on" until the neutral gas is illuminated along the light cone ($t_\mathrm{LC}\ga30$ Myr). Additionally, the lack of pre-heating by hard ionizing photons could reduce the emission somewhat. Thus, our results should be seen as optimistic.

There may be other significant sources of line photons at the very faint level of our prediction for the IF \lya signal. For example, scattering of non-ionizing quasar photons from the damping wing of the neutral IGM \citep{LR99} could be brighter than the IF \lya signal. In that case, a narrowband detection of extended emission would not necessarily indicate the presence of an IF, although one could still infer the presence of the neutral IGM. The Loeb-Rybicki halo emission would be much broader in velocity space ($\Delta v\ga1000$ km/s), so a spectral observation could in principle disentangle the two.

Analogous to the \HIa IFs considered here, one might also expect \HeIIa \lya emission from \HeIIa IFs during helium reionization at $z\sim3$. \HeIIa IFs are relatively broad ($dR_\IF\sim1$ pMpc; \citealt{FO08}) so in principle the IF emission could be brighter. However, \HeIIa resonance lines require four times as much energy to excite relative to \HIb, so the temperature within the IF is not high enough to produce any significant emission.

The detection of \lya emission from a quasar ionization front would be ``smoking gun" evidence of the reionization process (and potentially a useful diagnostic of quasar properties). Unfortunately, we have found that the largest existing telescopes fall far short of detecting our predicted signal. Even the next generation of thirty-meter class near-infrared telescopes will struggle mightily to detect these ionization fronts at any reasonable significance; measurements will require quasars with very hard spectra in relatively neutral environments and extremely long integrations.  The smaller background in space makes a satellite observation easier, but collecting area will likely be a major problem. Nevertheless, should the proper target quasar appear, this is an extremely powerful probe of the reionization epoch.

\section*{Acknowledgements}
We thank S. Cantalupo for comments on an earlier draft of the manuscript which motivated the discussion in Section \ref{sect:subgrid}. This work was partially supported by the David and Lucile Packard Foundation and by the NSF REU program at UCLA. MM acknowledges support by NSF grant AST-1312724.

\bibliographystyle{mn2e}

\end{document}